\documentclass[preprint2]{aastex62}

\usepackage{xcolor}

\usepackage{calrsfs,euscript,mathrsfs,amssymb}

\shorttitle{Galaxies at Gamma Rays}
\shortauthors{Ajello, Di Mauro, Paliya \& Garrappa}

\begin{document}


\title{The $\gamma$-ray Emission of Star-Forming Galaxies}


%
%
%
%

\author{M.~Ajello}
\email{majello@g.clemson.edu}
\affiliation{Department of Physics and Astronomy, Clemson University, Kinard Lab of Physics, Clemson, SC 29634-0978, USA}
\author{M.~Di~Mauro}
\email{mdimauro@slac.stanford.edu}
\affiliation{NASA Goddard Space Flight Center, Greenbelt, MD 20771, USA}
\affiliation{Catholic University of America, Department of Physics, Washington DC 20064, USA}
\author{V.~S.~Paliya}
\email{vaidehi.s.paliya@gmail.com}
\affiliation{Deutsches Elektronen Synchrotron DESY, D-15738 Zeuthen, Germany}
\author{S.~Garrappa}
\affiliation{Deutsches Elektronen Synchrotron DESY, D-15738 Zeuthen, Germany}

\begin{abstract}
A majority of the $\gamma$-ray emission from star-forming galaxies is generated by the
interaction of high-energy cosmic rays with the interstellar gas and radiation
fields. Star-forming galaxies are expected to contribute to both the
extragalactic $\gamma$-ray background and the IceCube astrophysical neutrino
flux. 
Using roughly 10\,years of $\gamma$-ray data taken by the {\it Fermi} Large Area
Telescope, in this study we constrain the
$\gamma$-ray  properties of star-forming galaxies. We report the detection
of 11 bona-fide $\gamma$-ray emitting galaxies and 2 candidates. Moreover, we show that
the cumulative $\gamma$-ray emission of
below-threshold galaxies is also significantly detected at
$\sim$5\,$\sigma$ confidence.
The $\gamma$-ray luminosity of resolved and unresolved galaxies  is
found to correlate with the total (8-1000\,$\mu$m) infrared  luminosity as
previously determined. 
Above 1\,GeV,  the spectral energy distribution
of resolved and unresolved galaxies is found to be compatible with a power
law with a photon index of $\approx2.2-2.3$.
 Finally, we find that star-forming galaxies account for roughly 5\,\% and 3\,\% of
the extragalactic $\gamma$-ray background and the IceCube neutrino
flux, respectively.
\end{abstract}


\section{Introduction}
The origin of the extragalactic $\gamma$-ray background (EGB) has been debated since its 
discovery by the OSO-3 satellite \citep{kraushaar73}. 
 The EGB spectrum has been
measured with good precision from 100\,MeV to 820\,GeV by
the Large Area Telescope (LAT) on board of the {\it Fermi 
Gamma-Ray Space Telescope} \citep{lat_egb2} and has been ascribed
to the emission of { resolved and} unresolved point sources
like blazars, star-forming and radio galaxies
\citep{ajello15,dimauro14b}. While the contribution of
blazars to the EGB is reasonably well constrained and understood
\citep[see, e.g.,][]{DiMauro:2013zfa,ajello15,ackermann16_egbres,dimauro18}, the
contribution from star-forming and radio galaxies is not. 
{ Analyses that have estimated the contribution of star-forming and
  radio galaxies to the EGB were based on 7 and 12 detected sources, respectively
  \citep[][]{lat_starforming,dimauro13}, yielding considerable uncertainties.}

Taking advantage of the increased sensitivity provided by ten year of
LAT data, in this study we re-evaluate the $\gamma$-ray properties of
star-forming galaxies (SFGs) whose
$\gamma$-ray emission is
{ partly of hadronic origin, being generated}
 by
cosmic-ray (CR) interactions with ambient gas and interstellar
radiation fields. The detection of
a flux of astrophysical neutrinos by IceCube \citep{icecube_nubkg13,PhysRevLett.113.101101,icecube_nubkg15}
with TeV-PeV energies has renewed the interest in 
SFGs since the
hadronic interactions that generate high-energy $\gamma$ rays inevitably
lead to the generation of high-energy neutrinos. 

The lack of 
anisotropy in the IceCube signal and template analyses show that most
($\gtrsim86$\,\%) of the IceCube signal is likely of extragalactic
origin \citep[][]{icecube_galactic17}. Moreover, the absence of
neutrino point-like sources and the non-detection of muon neutrino
multiplets at $>50$\,TeV imply that the population responsible for the majority of
the IceCube neutrinos is produced by unresolved sources with a  local density of
$\geq10^{-8}$\,Mpc$^{-3}$ or $\geq10^{-6}$\,Mpc$^{-3}$ for evolving
and non-evolving populations  respectively \citep{ahlers14,murase16,peng2017}.
Such populations could be jetted active galactic nuclei (blazars
and/or radio galaxies) and star-forming galaxies. However, it seems
unlikely that either of them can account for the totality of the
IceCube neutrino flux. Indeed, while a neutrino event has been found in
suggestive ($\approx$3\,$\sigma$) coincidence with  a flaring blazar  \citep{icecube18_txs}, cross-correlation studies
of all the neutrino events and LAT blazars demonstrated that blazars can
only produce a fraction $\lesssim27$\,\% of the neutrino intensity above
10 TeV \citep{icecube17_2lac}. Similarly, the non-blazar fraction
of the  EGB at $>50$\,GeV \citep{ackermann16_egbres} limits the contribution of SFGs
to the diffuse neutrino background to $\lesssim$30\,\% \citep{bechtol17}.

In this paper we use roughly 10 years of {\it Fermi}-LAT data to
study the global properties of SFGs and constrain
their contribution to the EGB and IceCube astrophysical   signal.
This paper is organized as follows:
$\S$~\ref{sec:sample} and $\S$~\ref{sec:analysis}  present the properties of the sample and method used
for the analysis. The results for the properties of SFGs are presented 
in  $\S$~\ref{sec:results}, $\S$~\ref{sec:corr}, and
$\S$~\ref{sec:spec}, and their contribution to the EGB and IceCube
neutrino flux is discussed in $\S$~\ref{sec:EGBnu}. Finally,
$\S$~\ref{sec:conclusions} summarizes the results.
Throughout this paper, a standard concordance cosmology was assumed
(H$_0$=71\,km s$^{-1}$ Mpc$^{-1}$, $\Omega_M$=1-$\Omega_{\Lambda}$=0.27).

%
%
%
%
\section{Sample of Galaxies}

\label{sec:sample}
Massive star formation is fueled by  dense molecular gas in the interstellar medium (ISM). 
In order to select a sample of galaxies with unambiguous ongoing star formation, we base our sample of galaxies on the  following resources:
\begin{itemize}
\item The survey of HCN emission from 65 infrared (IR) galaxies published in \cite{2004ApJ...606..271G}.
This includes nine ultra-luminous infrared galaxies, 22 luminous infrared galaxies, and 34 normal spiral galaxies with low IR luminosity.
\item The sample of 83 nearby starburst galaxies with UV-to-FIR data
  selected based on the availability of UV data from the {\it International
  Ultraviolet (UV) Explorer} \citep{2002ApJS..143..377W}.
This sample was also used in the previous {\it Fermi}-LAT
analysis of SFGs \cite[][hereafter ACK12]{lat_starforming}.
\item The {\it Infrared Astronomical Satellite} (IRAS) Revised Bright Galaxy Sample (RBGS).
This is a complete flux-limited survey of all extragalactic objects with total 60 micron flux density greater than 5.24 Jy.
This sample covers the entire sky surveyed by {\it IRAS} at Galactic latitude $|b| > 5^{\circ}$. 
The RBGS includes 629 objects, with a median  sample redshift of $z=$0.0082 and a maximum redshift of $z=$0.0876 \citep{Sanders:2003ms}.
\end{itemize}

The selection of the above catalogs results in a sample of 683
SFGs with known position, redshift, distance,
and infrared luminosity (L$_{\rm IR}$) in the wavelength range
$8-1000$\,$\mu$m. L$_{\rm IR}$ is derived by fitting a single temperature
dust emissivity model to the flux reported in the 4 {\it IRAS} bands and
should be accurate to $\pm$5\% for dust temperatures in the 25-65\,K
range \citep[see][for details]{sanders96}.

{ We then remove all blazars from our list based on positional
  coincidence with sources from the following two catalogs. First, we
  remove all sources that are positionally coincident with any of the
  associated point sources (other than associations with galaxies)
  from the preliminary \textit{Fermi}-LAT 8-year point source list
  (FL8Y\footnote{This worked started well before the LAT 8-year source
    catalog \citep[4FGL][]{4fgl} became available. For the FL8Y see:
  \url{https://fermi.gsfc.nasa.gov/ssc/data/access/lat/fl8y}}). Additionally,
we remove all sources that are positionally coincident with sources
from the 5th edition of Roma-BZCAT \citep{bzcat5}, which is one of the most up-to-date lists of blazars. These cuts reduce the number of SFGs in our sample to 588, which will be considered in the rest of our analysis.  }

\begin{figure*}[ht!]
  \begin{center}
  \begin{tabular}{c}
\hspace{-1cm}
  	 \includegraphics[scale=0.65,clip=true,trim=0 0 0
    0]{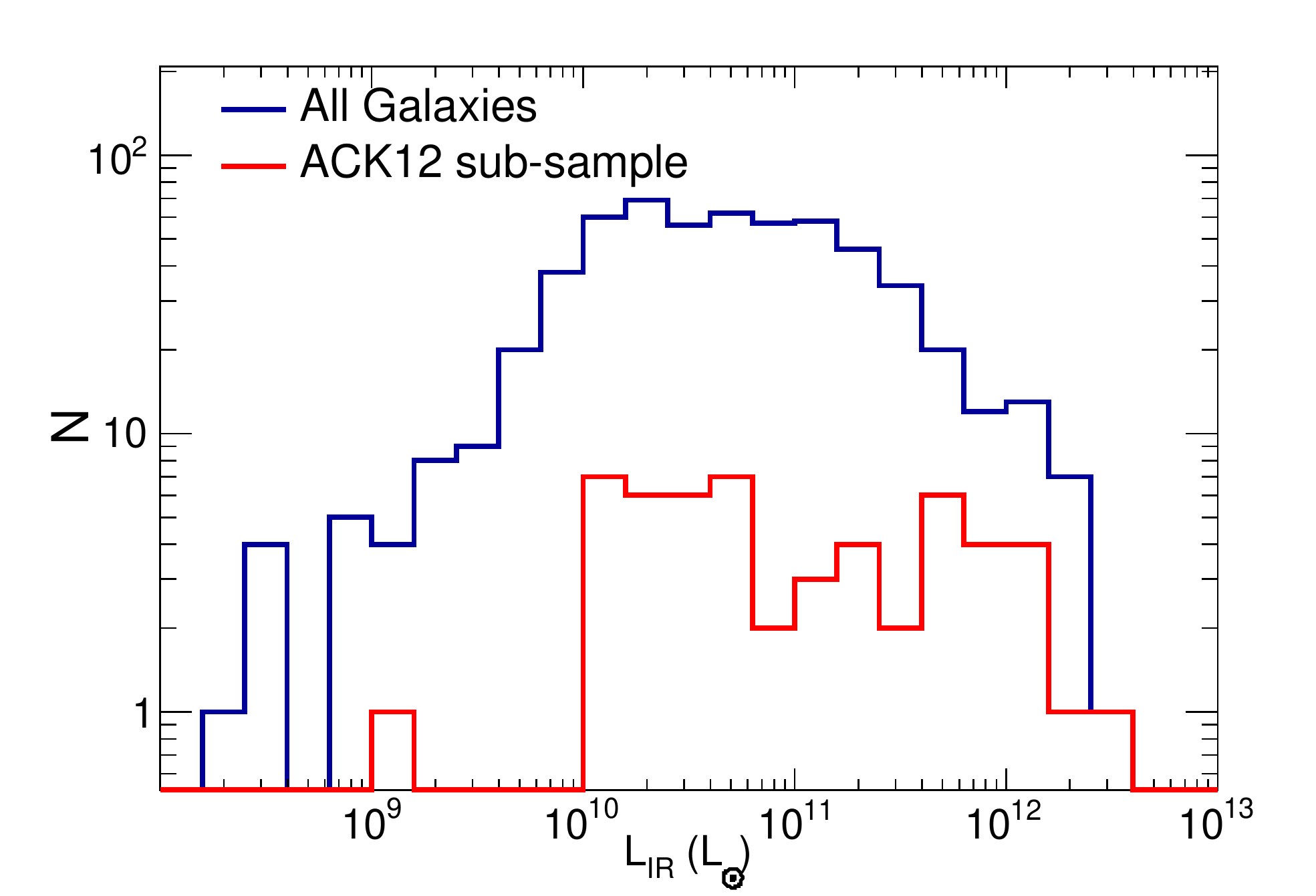} 
\end{tabular}
  \end{center}
  \caption{Distribution of infrared luminosity for the full sample of galaxies described in
    $\S$~\ref{sec:sample} and the subset of objects of ACK12.
\label{fig:lir_dist}}
\end{figure*}

In Figure~\ref{fig:lir_dist} we show the distribution of the IR
luminosity, L$_{IR}$, for our full sample of SFGs and the subset of
galaxies used in ACK12.
The peak of the distribution is around $Log (\rm{L}_{\rm IR} (L_{\sun}))=10-11$, but
the full sample contains many more galaxies with L$_{\rm IR}< 10^{10}$
$L_{\odot}$ compared to the ACK12 subset.

\section{{\it Fermi}-LAT Data Analysis}
\label{sec:analysis}

%
%
\subsection{Analysis of Individual Galaxies}
\label{sec:analaysis_det}
We analyze { the} {\it {Fermi}}-LAT data for each source in our sample in
order to  { determine} whether { or not} it is detected (with a
$TS$\footnote{The Test Statistic, $TS$, is defined as $TS=2(\log L_1
  - \log L_0)$, where $\log L_1$ and $\log L_0$ are the
  log-likelihood values for the case of  background plus source and
  only background respectively \citep{mattox96}.}$>25$).
We analyze almost 10 years of Pass 8 data, from 2008 August 4 to 2018
March 4, selecting $\gamma$-ray events in the energy range
$E=[0.1,800]$ GeV, passing the standard data quality selection
criteria\footnote{We apply the filter $(DATA\_QUAL>0)\&\&(LAT\_CONFIG==1)$ with the {\tt gtmktime} tool of the Science Tools. See \url{https://fermi.gsfc.nasa.gov/ssc/data/analysis/scitools/binned_likelihood_tutorial.html} for further details.}.
We consider events belonging to the Pass~8 {\tt SOURCE} event class,
and use the corresponding  instrument response functions (IRFs) {\tt
  P8R3\_SOURCE\_V2}, since we are interested in  the detection and
study of point sources.

We reduce the contamination from the low-energy Earth limb emission by applying a zenith angle cut to the data.
Following { the}
FL8Y\footnote{\url{https://fermi.gsfc.nasa.gov/ssc/data/access/lat/fl8y/FL8Y\_description\_v5.pdf}}
{ analysis},
we { make} a harder cut at low energies { by selecting} event types with the best
point spread function (PSF). For $E=[0.1,0.3]$\,GeV, we select events
belonging to the PSF2 and PSF3 classes with zenith angles 
$<$90$^{\circ}$. For $E=[0.3,1.0]$\,GeV  { we select} PSF1, PSF2 and PSF3 event types
with zenith angles $<$100$^{\circ}$. And above 1
GeV we use  all events (PSF0, PSF1, PSF2, PSF3) with zenith angles less than 105$^{\circ}$.
This choice reduces the contribution of the Earth limb contamination
to less than $10\%$ of the total background.

We employ an analysis pipeline based on {\tt FermiPy}, { which is}
a Python package, based on the {\it Fermi} Science Tools, that automates analyses \citep{2017arXiv170709551W}\footnote{See \url{http://fermipy.readthedocs.io/en/latest/}.}.
{\tt FermiPy} includes tools that perform fits of the model to the
$\gamma$-ray data, detect new point sources, calculate the source
spectral energy distribution (SED), { and} test for the presence of source extension and variability.   

{ For each galaxy in our sample we consider a region of interest (ROI)
of $12^\circ \times 12^\circ$ centered at the respective galaxy, and
each ROI is analyzed separately. }
In each ROI, we bin the data with a pixel size of $0.06^{\circ}$ and 8 energy bins per decade.
We model the Galactic diffuse emission using the interstellar emission model (IEM)
released with Pass 8 data \citep[{\tt gll\_iem\_v06.fits,} ][]{Acero:2016qlg}.
{ Correspondingly, we use} the standard template for the isotropic
emission ({\tt iso\_P8R2\_SOURCE\_V6\_v06.txt})~\footnote{For
  descriptions of these templates, see
  \url{http://fermi.gsfc.nasa.gov/ssc/data/access/lat/BackgroundModels.html}.}.
{Two objects, the Large and the Small Magellanic clouds (LMC and
  SMC), are modeled as spatially extended sources adopting the
  templates used in 
  the FL8Y (see App.~\ref{sec:appA}).}
The point sources are modeled using the FL8Y catalog.
{Energy dispersion  is applied to all sources in the model except the isotropic template\footnote{The
    isotropic template, which has been derived from observed data,
    includes already the effect of energy dispersion and does not need
  to be corrected.}. This includes the Galactic diffuse templates.
We
  used the implementation of the energy dispersion present in the
  Fermi tools\footnote{For more information see \url{https://fermi.gsfc.nasa.gov/ssc/data/analysis/documentation/Pass8_edisp_usage.html}}. Specifically we have fixed apply\_edisp=True and edisp\_bins$=-$1.}

During the optimization of  { each} ROI, 
we first relocalize the source of interest. { We} then search for new
point sources with $TS > 25$, which are inserted in the model. These
sources are found generating a $TS$ map where a test source is moved
across the ROI.
After this first step we test whether the emission of the SFG
candidate is consistent with
that of a point-like or spatially extended source.
 Out of the 588 galaxies searched in this way, 13 SFGs are 
detected with significance\footnote{The conversion between $TS$ and
  $\sigma$ was performed using a $\chi^2$ distribution with 2 degrees
  of freedom, corresponding to the two free spectral parameters.} $>4.6$\,$\sigma$  ($TS>$25). 
 Another galaxy (UGC~11041) is also 
detected but we later show that its emission is likely due to a
relativistic jet.
{ Properties of the detected galaxies}, including the SEDs and lightcurves, are
described in $\S$~\ref{sec:results}.

%
%
\subsection{Stacking Analysis of Unresolved Galaxies}
\label{sec:stacking}

Here we test whether the LAT detects the collective emission
from  galaxies too faint to be detected individually. 
We { use}  the sample of 575 undetected objects
 { from} the previous section. 
A pre-analysis is performed similar to the analysis of detected
objects (see Section~\ref{sec:analaysis_det}) with a few key
differences. First, we select photons of all
PSF types (0,1,2, and 3) from 0.1\,GeV to 800\,GeV within 
15$^{\circ}$ of the galaxy of interest to properly model the background. 
Similarly to $\S$~\ref{sec:analaysis_det}, background sources include those in the FL8Y
source list plus any new source detected here with a $TS>$25. These
sources are found with an iterative procedure generating a $TS$ map
and looking for excesses above the $TS>$25 threshold. 
 In order to retain sensitivity to a range of power-law spectral slopes,
$TS$ maps are generated for photon indices of 2.0, 2.5, and 3.0.
Once all excesses above background have been 
inserted in the background model a final fit is performed to optimize
all the free parameters { (those of the diffuse templates and background
sources)} in the ROI.

In order to detect the collective emission of { the} undetected
galaxies, we perform a stacking analysis where we sum the
log-likelihood profiles generated for the { spectral parameters of
  the SFGs} under the assumption { that the source population can
  be characterized by some average quantities (i.e., spectral index and
flux), as described in the next section}. In
generating the log-likelihood profiles the parameters of the isotropic
and Galactic diffuse templates are { left} free to vary, while all those of the background sources
are kept fixed at their optimized values. Further details on the
stacking are also reported in \cite{paliya2019}.

%
%
\section{Detection of Individual and Unresolved Galaxies}
\label{sec:results}

\begin{deluxetable*}{lccccccc}
\footnotesize
\tablecaption{List of SFGs detected in our analysis. { In this table,
  we report name}, coordinates, distance from Earth, IR luminosity,
  $TS$, $\gamma$-ray energy flux and luminosity of each source. The
  table is divided into two blocks. The upper block lists the bona fide
  sample, while the lower block reports the SFGs that are offset with
  respect to the IR-$\gamma$-ray luminosity correlation.
\label{tab:detected}}
\tablehead{
\colhead{Name} & \colhead{R.A. (J2000)}  & \colhead{Decl. (J2000)}  & \colhead{dist.\tablenotemark{a}}   & \colhead{$\log_{10}{L^{8-1000 \mu \rm{m}}_{\rm{IR}}}$}     & \colhead{$TS$}  & \colhead{$S$\tablenotemark{b}}  &  \colhead{$\log_{10}{L_{\gamma}}$}      \\
   \colhead{}        & \colhead{[deg]}  & \colhead{[deg]}      & \colhead{[Mpc]}  & \colhead{[$L_{\odot}$]}
           & \colhead{}           & \colhead{} & \colhead{[erg s$^{-1}$]}   
}
\startdata
M31  &  $10.69$ & $41.27$  &  0.78  &  9.39  &  122  &   $0.63\pm0.07$  & 38.66 \\
NGC 253  &  $11.89$ & $-25.28$  &  3.30 &  10.44  &  608  &   $0.89\pm0.07$  & 40.05 \\
SMC  &  $13.18$ & $-72.83$  &  0.063  &  7.86  &  1938  &  $2.90\pm0.12$  & 37.14\\
M33  &  $23.47$ & $30.67$  &  0.85  &  9.07  &  41 &   $0.21\pm0.04$  & 38.25 \\
NGC 1068  &  $40.68$ & $-0.01$  &  10.1  &  11.27  &  335  &  $0.68\pm0.05$   & 40.92 \\
LMC  &  $80.89$ & $-69.76$  &  0.05  &  8.82  &  4797  &  $10.7\pm0.3$  & 37.50 \\
NGC 2146  &  $94.66$ & $78.36$  &  18.0  &  11.07  &  73  & $0.23\pm0.04$  & 40.95\\
M82  &  $148.97$ & $69.68$  &  3.70 &  10.77  &  1186  &   $1.13\pm0.05$  &  40.27   \\
Arp 299  &  $172.13$ & $58.57$  & 47.7  &  11.88  &  32  &   $0.13\pm0.04$  & 41.55\\
NGC 4945  &  $196.36$ & $-49.47$  &  3.80  &  10.48  &  489  &   $1.15\pm0.07$  & 40.30\\
Arp 220  &  $233.74$ & $23.51$  &  79.9  &  12.21  &  84  &   $0.35\pm0.05$  & 42.42 \\
\hline
\hline
NGC 2403  &  $114.22$ & $65.61$  &  3.2  &  9.19  &  49  &  $0.17\pm0.04$  & 39.33 \\
NGC 3424  &  $162.94$ & $32.90$  &  27.2  &  10.30  &  40  &   $0.17\pm0.05$  & 41.18 \\
\enddata
\tablenotetext{a}{{Distances are
  from: ACK12 for the Local Group galaxies (M31, SMC, M33, and LMC),
\cite{mouhcine2005} for NGC~253,  
 \cite{nasonova2011} for  NGC~1068, 
\cite{adamo2012} for NGC~2146,
\cite{vacca2015} for M82,
the redshift for Arp 229 and Arp 220,
\cite{mould2008} for NGC 4945,
\cite{tully2013} for NGC 2403, and
\cite{theureau2007} for NGC 3424.
IR luminosities are taken from
\cite{2002ApJS..143..377W},  \cite{Sanders:2003ms}, and \cite{2004ApJ...606..271G}.}}
\tablenotetext{b}{$\gamma$-ray flux in units of $10^{-11}$ erg cm$^{-2}$
             s$^{-1}$.}
\end{deluxetable*}

\subsection{Individual Galaxies}
\label{sec:detected}
{ From our analysis we} detect significant $\gamma$-ray
emission ($TS>25$) at the positions of 13 SFGs, whose characteristics are reported in
Table~\ref{tab:detected}. Our sample of detected galaxies contains the
seven already reported in ACK12 (LMC, SMC, M31, M82, NGC~253, NGC~4945
and NGC~1068). In addition to those,  NGC~2146 and Arp~220
\citep[reported already by][]{Tang:2014dia,2041-8205-821-2-L20} are
also detected in this work. Gamma-ray emission from
NGC~2403 and NGC~3424 has also been previously reported by
\cite{linden_2017}, who, however, flagged them as potential background
fluctuations. Finally, our analysis reports the { first} detection of
emission from M33\footnote{We note that a source is
  detected at the position of M33 also in FL8Y. However, such source is not
present in 4FGL possibly due to the updated diffuse model. Its effect
will be studied in  a future publication.} and  Arp 299. Note that we also detect
significant ($TS=44$) $\gamma$-ray emission from UGC
11041, but {  due to its strong variability (see
  Table~\ref{tab:seddetected} and App.~\ref{sec:lc}), this emission is likely not
  associated to star-formation processes but to a 
  jet. Therefore  UGC~11041 is not part of our sample of
detected galaxies.}

M33, also known as the Triangulum galaxy, 
{ at  a distance of ~870 kpc,  is the smallest
  spiral galaxy in the Local Group. Its star-formation rate (SFR) is in the range 
0.26-0.70 \,M$_{\odot}$ yr$^{-1}$ \citep{2007A&A...473...91G}.}

Arp 299 at $\sim48$\,MpC, is much more distant than
M33 and it is 
an interacting system, at an early dynamical stage, composed of two individual galaxies.
Powerful starburst regions with SFRs of about 100
\,M$_{\odot}$ yr$^{-1}$ have been identified there by
\cite{2000ApJ...532..845A}. 
We consider all $\gamma$-ray sources  except NGC~3424 and NGC~2403 as part of
our bona-fide sample of galaxies. These two galaxies will be discussed in
more detail in $\S$~\ref{sec:outliers}.

\begin{figure*}
\includegraphics[scale=0.42]{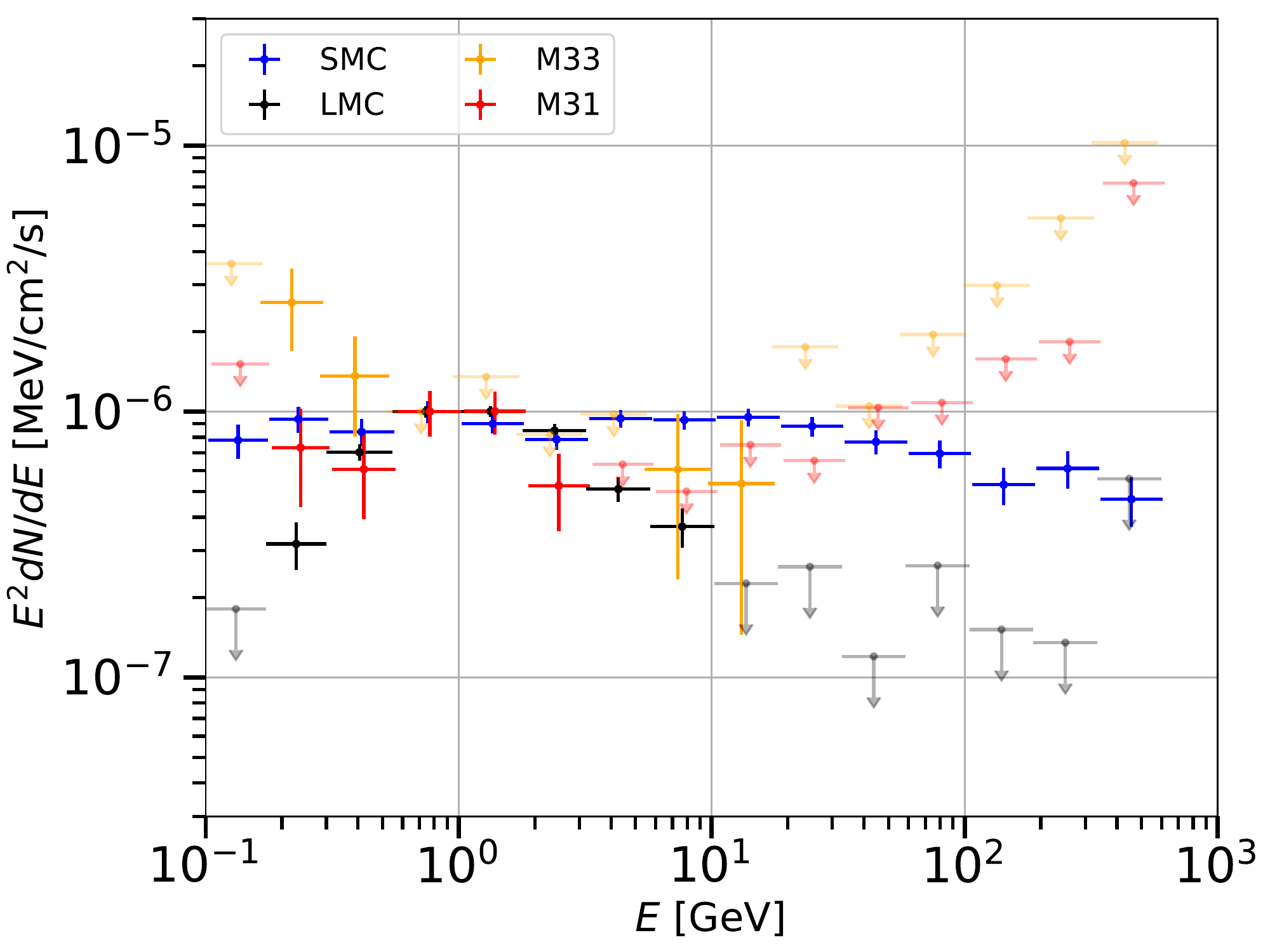}
\includegraphics[scale=0.42]{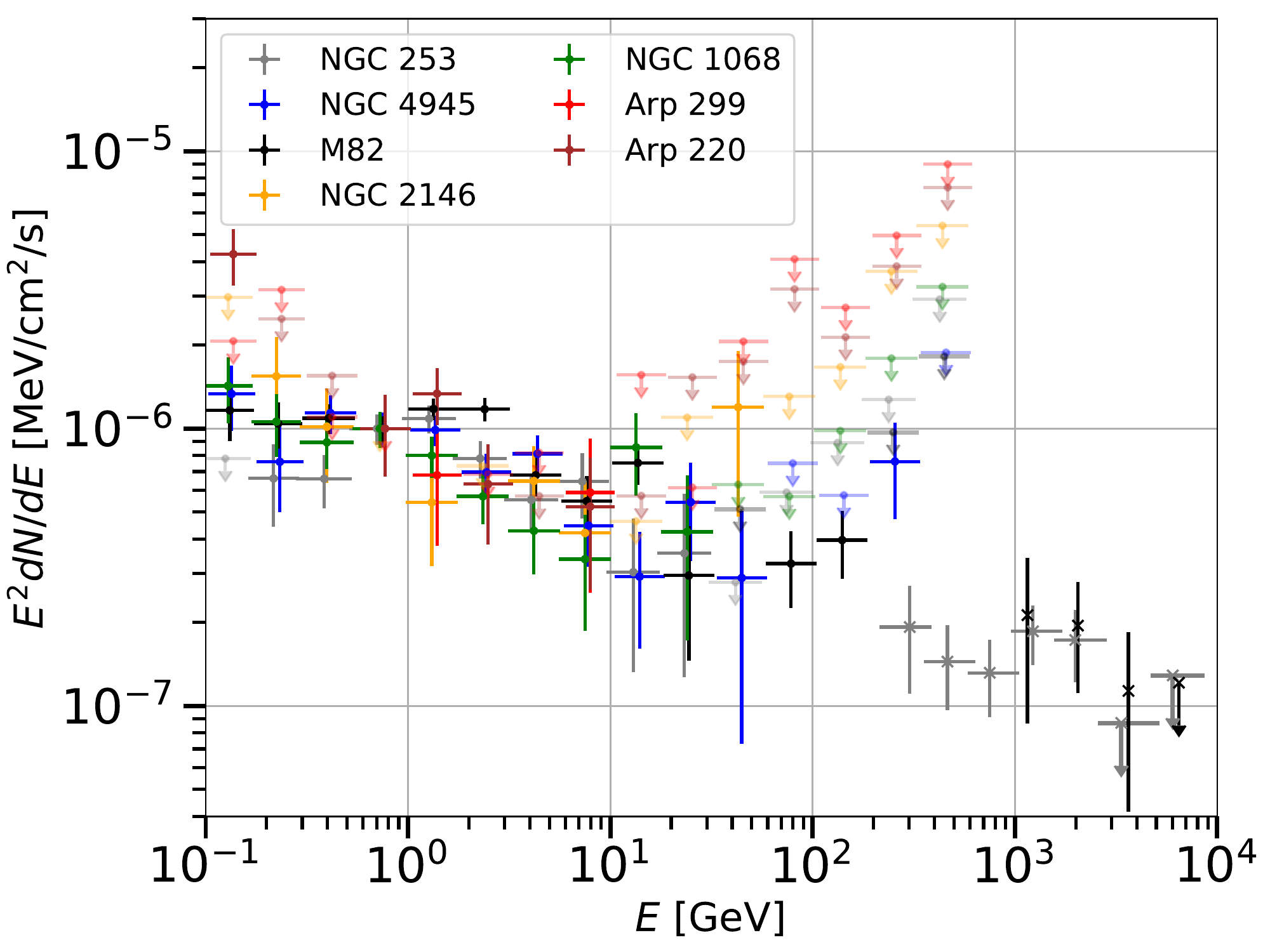}
  \caption{SEDs of bona fide SFGs. The
    galaxies have been ordered by increasing  IR luminosity. Left Panel:
    SEDs for SMC, LMC, M33, and M31. Right Panel: NGC 253, NGC 4945,
    M82, NGC 2146, NGC 1068, Arp 299, and Arp 220. 
    For NGC 253 and M82 
    we have included the SEDs { measured} by H.E.S.S. \citep{Abdalla:2018nlz} and 
    VERITAS 
    \citep{2009Natur.462..770V} (crosses), respectively. Upper limits
    ({for $TS<6$}) are plotted with 
    lighter colors for visualization purposes.}
\label{fig:sed_all1}
\end{figure*}

\begin{figure}
\includegraphics[scale=0.42]{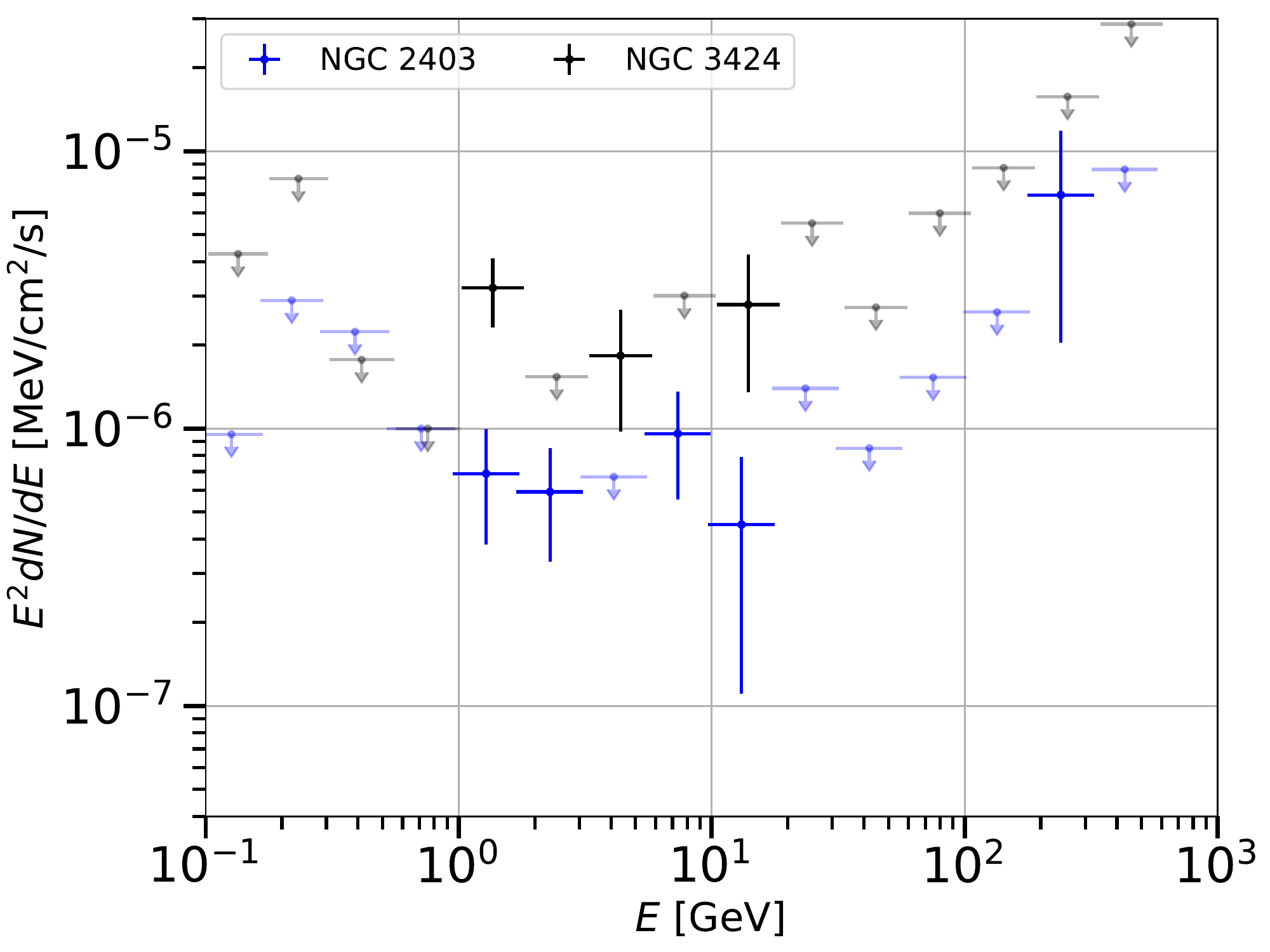}
  \caption{SEDs for candidate SFGs: NGC 2403 and  NGC
    3424.  { Upper limits ({for $TS<6$})  are plotted with 
    lighter colors for visualization purposes.}}
\label{fig:sed_all2}
\end{figure}

We test whether the $\gamma$-ray spectra of the detected SFGs
{ are} described better
by a power law (PL) or a log-parabola (LP). 
These SED shapes are defined as follows:
 \begin{equation}
 \frac{dN}{dE} = K \left( \frac{E}{E_0} \right)^{-\Gamma},
 \label{eq:PL}
 \end{equation}
for the PL, where $K$ is the normalization, $E_0$ is fixed to 1\,GeV, and $\Gamma$ is the spectral index;
 \begin{equation}
 \frac{dN}{dE} = K \left( \frac{E}{E_0} \right)^{-\Gamma + \beta \log{(E/E0)}},
 \label{eq:LP1}
 \end{equation}
for the LP, where $\beta$ is the curvature index.
In order to choose between the above models we compute the $TS$ of
curvature ($TS_{\rm{curv}}$) defined as $TS_{\rm{curv}} = -2 ( \log{L_{\rm{PL}}} - \log{L_{\rm{LP}}} )$.
Only two galaxies, { the} LMC and the SMC, have a significant
($TS_{\rm{curv}}>25$, i.e. $>$5\,$\sigma$) curvature. The SED
parameters of { all }
these { detected} galaxies are reported in Table~\ref{tab:seddetected}, while their
SEDs are { plotted} in Figures~\ref{fig:sed_all1} and \ref{fig:sed_all2}.

Finally, we extract a light curve for each galaxy using
the tools available in {\tt Fermipy}.
This is { done} by dividing the entire dataset into time intervals 
of { one}-month duration for the six brightest sources and 3 month duration
for the rest.  Having more bins increases the ability to detect
variability for bright sources.  On the other hand, for fainter sources one-month time bins are too narrow to allow detection of the source in most bins.
For each time interval a fit of the model to the data is performed,
{ from which we obtain}  the flux, SED,
and detection $TS$.
In the final step the $TS$ for the variability ($TS_{\rm{var}}$, see
Appendix~\ref{sec:lc}) is
calculated following \cite{2FGL}.
Since,  different time bins are used for different sources, we report
in Table~\ref{tab:seddetected} the values of the significance of
variability $\sigma_{\rm{var}}$, computed considering that there are 119 time bins and 118 degrees of freedom for the first six bright
galaxies and 39 time bins and  38 degrees of freedom for the rest.

As already mentioned, UGC 11041 is the only galaxy with a significant variability (5.7\,$\sigma_{\rm{var}}$). 
Moreover, this source underwent a
strong flare between MJD 56600 and 56800, { as can be
  observed from its lightcurve (see App.~\ref{sec:appA})}. 
This flare, together with the spatial coincidence
of the galaxy with a radio source in the NVSS catalog,
{ suggests that the observed $\gamma$-ray emission is due to a relativistic jet, as opposed to star-formation activity.}
This source is thus excluded from further analysis.
The lightcurves of all the galaxies that we detected are shown in the Appendix
$\S$~\ref{sec:lc}.

\begin{table*}[t]
\center
\begin{tabular}{lccccc}
\hline 
Name & $TS_{\rm{curv}}$   &  $\Gamma$   &   $\beta$   &  $E_0$  & $\sigma_{\rm{var}}$   \\
           &				&      			&   & 	[GeV] &	  \\
\hline 
\hline
LMC  & 335  & 1.42$\pm0.05$  & -0.42$\pm0.04$   & 2.60$\pm0.35$  &1.4    \\
SMC  &  65  & 1.73$\pm0.06$  & -0.10$\pm0.03$   & 1.25$\pm0.18$  &2.9  \\
M82  & 13  &  2.14$\pm$0.06 &    &  & 2.0 \\
NGC 253  &  13 &  2.10$\pm$0.06 &   &   & 1.1 \\
NGC 4945  & 2.3  & 2.22$\pm$0.07  &    &   & 1.1   \\
NGC 1068  & 0.0  & 2.27$\pm$0.09  &    &   & 0.1  \\
M31  & 6.0  & 2.23$\pm$0.11  &    &    & 0.7 \\
Arp 220  & 1.3  & 2.48$\pm$0.14   &     &    & 0.0 \\
M33  & 0.4  & 2.41$\pm$0.16  &    &   & 0.5  \\
NGC 2146  & 1.3  & 2.27$\pm$0.07  &    &     & 0.9\\
Arp 299  & 0.1  & 2.11$\pm0.19$  &    &     & 0.1\\
\hline
\hline
NGC 2403  & 0.0  & 1.94$\pm0.19$  &    &   & 0.7  \\
NGC 3424  & 2.5  & 1.97$\pm0.20$  &    &    & 0.7 \\
\hline
UGC 11041  & 6.9  & 2.53$\pm0.22$  &    &   & 5.7  \\
\hline 
\end{tabular}
\caption{SED parameters and significance of variability ($\sigma_{\rm var}$) for the SFGs detected in our analysis. We
  report the $TS$ for the curvature ($TS_{\rm{curv}}$) and the SED
  parameters for the power-law model if $TS_{\rm{curv}}<25$ and for
  the 
  logparabola model if $TS_{\rm{curv}}\geq 25$. The last column
  reports the significance of variability
variability ($\sigma_{\rm{var}}$, see the text and Appendix for further details).
\label{tab:seddetected}}
\end{table*}

%
%
\subsection{Unresolved Galaxies}
\label{sec:flux_stack}
\begin{figure*}[ht!]
  \begin{center}
  \begin{tabular}{ll}
\hspace{-1cm}
  	 \includegraphics[scale=0.40,clip=true,trim=0 0 0
    0]{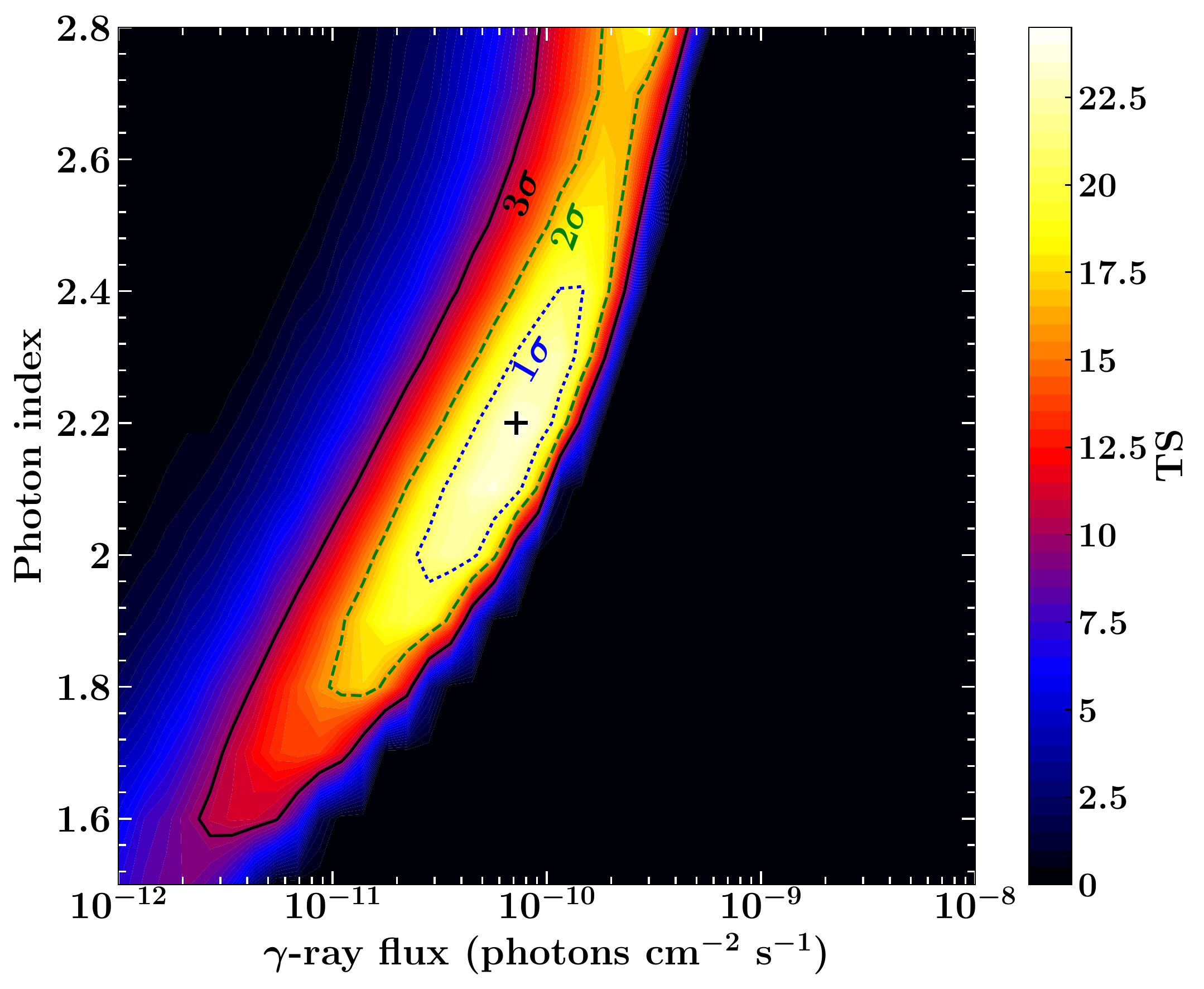} &
\hspace{-0.2cm}
	 \includegraphics[scale=0.40,clip=true,trim=0 0 0
    0]{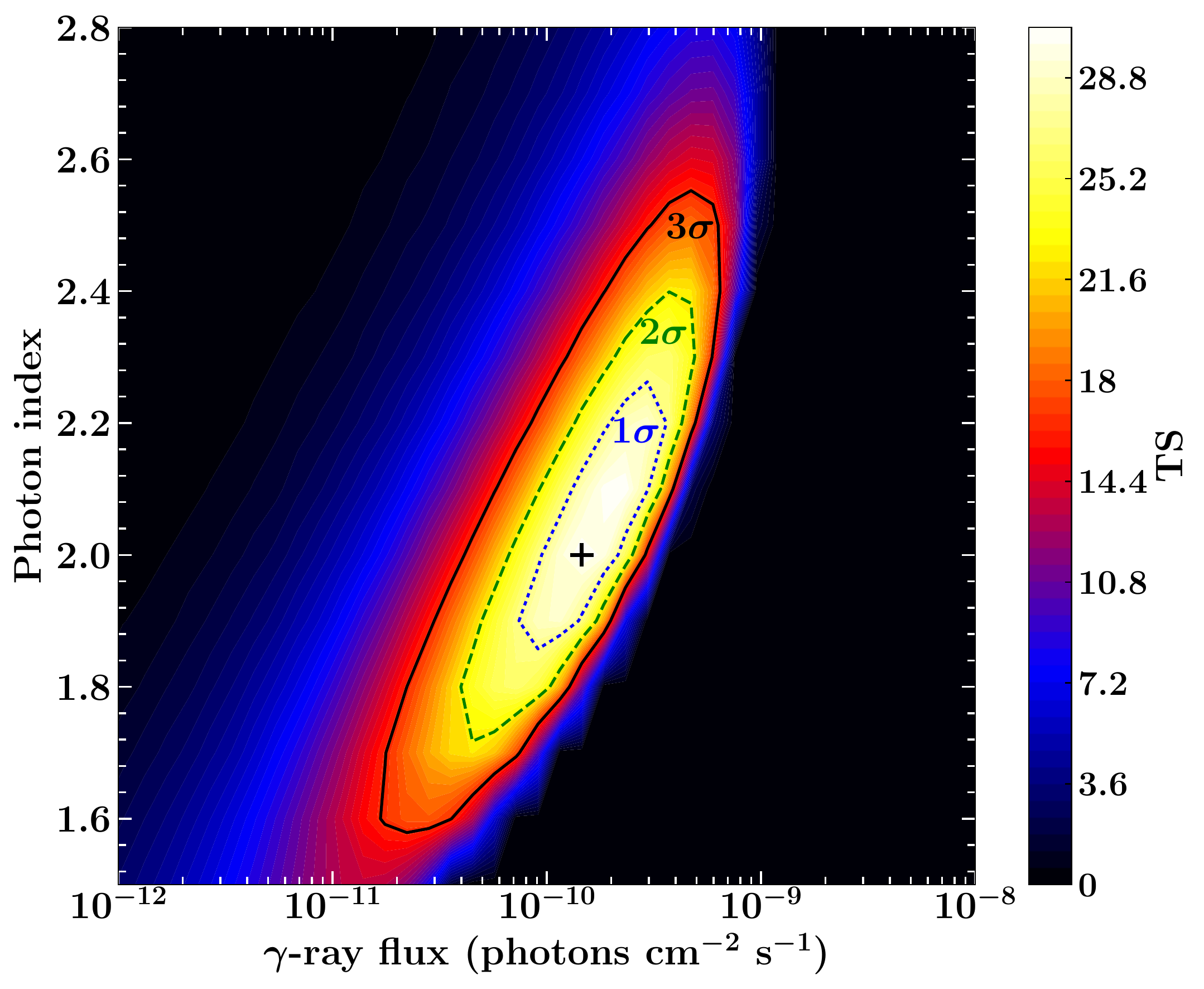} \\
\end{tabular}
  \end{center}
  \caption{Detection significance as a function of photon index and
    $\gamma$-ray flux for the full sample of 574 undetected galaxies (left) and
    for the 56 undetected galaxies (right) in ACK12 sub-sample. 
\label{fig:flux_stack}}
\end{figure*}

For each undetected galaxy, whose spectrum is modeled { with} a power law, we generate a bi-dimensional log-likelihood profile varying its photon index and photon flux (measured in the 100\,MeV --
800\,GeV energy range). 
By subtracting   the log-likelihood of the null hypothesis ($\log
L_{null}$), i.e., the likelihood of the data if the galaxy in question
is not a gamma-ray emitter,  
 we define the $TS=2(\log L - \log
L_{null})$. We sum the $TS$ profiles of every galaxy to quantify
the significance of their collective detection, assuming that the
entire population can be characterized by an average photon index and
an average flux. Figure~\ref{fig:flux_stack} shows the results of
{ this stacking analysis}
 for the entire sample of undetected
galaxies (574, left) and the sub-sample of undetected galaxies
reported in \cite{lat_starforming} (56, right).

Figure~\ref{fig:flux_stack} { illustrates} that
the average emission from all unresolved galaxies is detected by the LAT
with a $TS$$=$23 (corresponding to $\sim$4.4\,$\sigma$). 
The average $\gamma$-ray emission of the ACK12 sub-sample is 
detected with $TS$$=$30 ($\sim$5.1$\sigma$).
The signal is
stronger for the ACK12 sub-sample { compared to} the entire sample (which
is $>10$ times larger in number) because the galaxies in ACK12 are distributed over
a narrower range of bright IR luminosities and fluxes (see Figure~\ref{fig:lir_dist}).
The average photon index for the two sets is $\Gamma=2.21\pm0.12$ and
$\Gamma=2.00\pm0.12$, respectively. The average $\gamma$-ray spectrum of unresolved SFGs is thus as hard as the typical spectrum of starburst
galaxies with L$_{\rm IR}=$10$^{11-12}$\,L$_{\odot}$
\citep[see Table~\ref{tab:seddetected} and][]{lat_starforming}.

%
%

\section{The L$_{\gamma}$-L$_{\rm IR}$ Correlation}
\label{sec:corr}

\subsection{Individual Galaxies}
\label{sec:corr_det}

\begin{figure*}
\centering
\includegraphics[width=0.7 \textwidth]{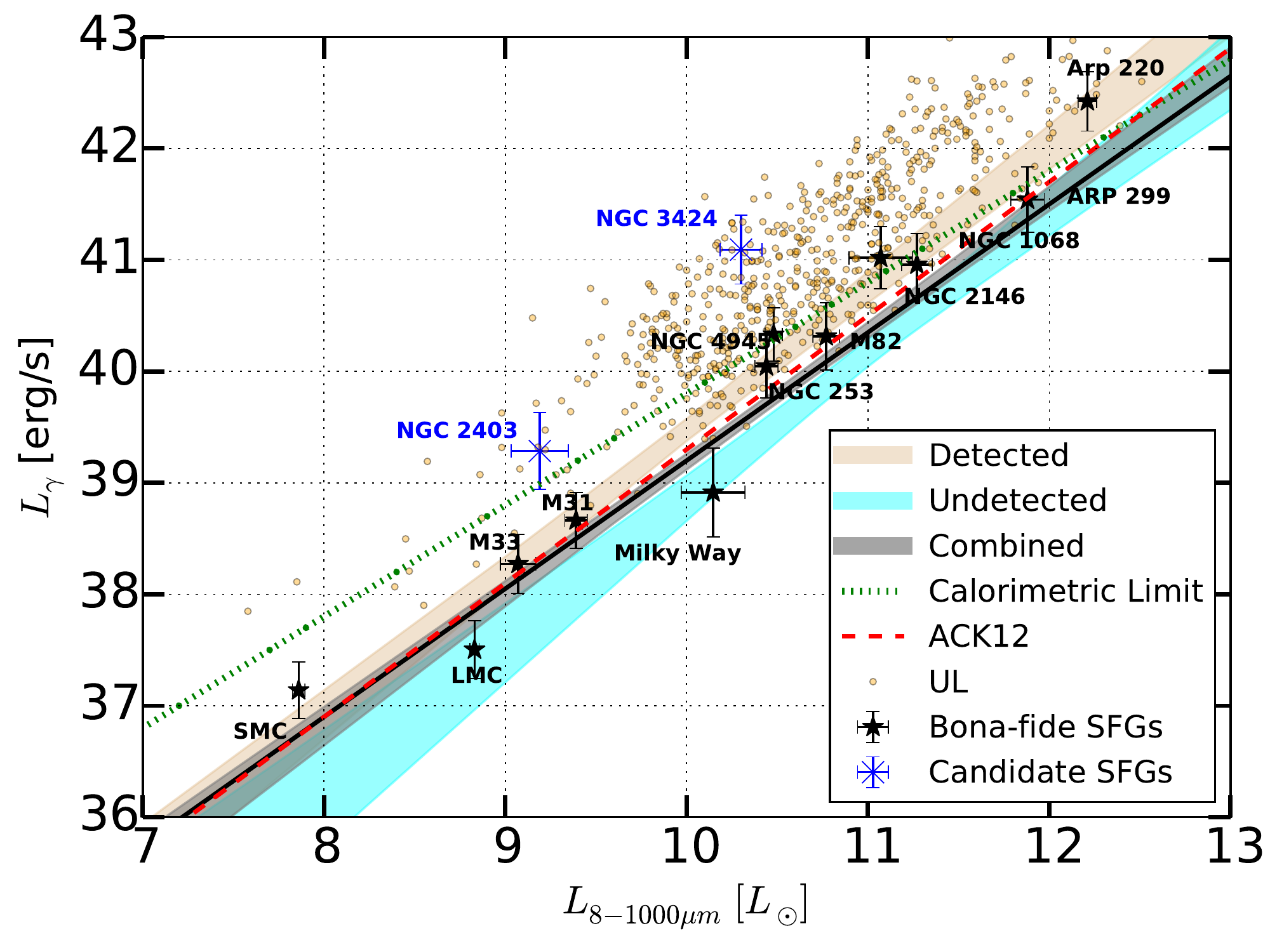}
\caption{Gamma-ray vs. IR luminosities for the SFGs in our sample
best fit. Bona fide SFGs are reported with black data points while galaxy
  candidates { are shown} with blue points. The best-fit 1\,$\sigma$ uncertainty band
  for the linear function in log space { of} the bona-fide sample is shown
  with a brown band, while the cyan band displays the  1\,$\sigma$  uncertainty band of the
  unresolved sources. The combined fit and its 1\,$\sigma$  uncertainty are plotted with the 
black line and the gray band, respectively.
{ For comparison the red dashed line shows}
the best-fit correlation published
in \cite{lat_starforming}. The brown data points show $\gamma$-ray {
  luminosity} upper limits  
for all the individual galaxies in our sample, {which are
  undetected in our analysis ($TS<25$)}.}
\label{fig:corrdetected}
\end{figure*}

As already shown in \cite{lat_starforming}, { for SFGs} a
correlation exists between the
$\gamma$-ray luminosity and the total IR luminosity, since both
quantities are related to star-formation activity.
Figure~\ref{fig:corrdetected} shows the relation between $\gamma$-ray and IR luminosities for the
objects detected in this work.
NGC
2403 and NGC 3424 have an offset with respect to the
general trend of all other SFGs: i.e., they are brighter in
$\gamma$ rays for a given IR luminosity.
Therefore, we decide to select LMC, SMC, M82, NGC~253, NGC 4945,
NGC 1068, M31, NGC 2146, M33, Arp 220, and Arp 299 as bona fide
sources (i.e., galaxies whose $\gamma$-ray emission can be confidently
attributed to star formation activity).

{ We fit a linear function (in log space)} to the $\gamma$-ray and IR luminosities, for this bona fide
  sample.
This
gives values of $\alpha = 1.27 \pm 0.03$ and $\beta = 39.38 \pm 0.03$,
{ where the function being  fit is defined as}:
 \begin{equation}
 \log_{10}{\left( \frac{L_{\gamma}}{\rm{erg/s}} \right) } = \beta + \alpha \log_{10}{\left( \frac{L_{\rm{IR}}}{ 10^{10} L_{\odot}} \right) }.
 \label{eq:corr}
\end{equation}
However, { the fit} results in a poor { reduced} chi-square
$\chi^2/d.o.f=175/11$. This is likely due to uncertainties in the distance of these galaxies
and scatter in their $\gamma$-ray and IR properties.
Indeed, within 100\,Mpc, the typical distance
is uncertain by $\sim$10-20\,\% \citep{freedman2001}.
The total IR luminosity (L$_{\rm IR}$) is also  { only} accurate within 5\,\%
\citep{sanders96}.
Finally, for dust-poor systems,  L$_{\rm IR}$ is not a
reliable tracer of star formation \citep[see e.g.,][]{hayashida2013}.
{ It is therefore not very surprising to find such}
scatter 
 in the
$L_{\gamma}-L_{\rm IR}$ relation.

{ Although quantifying these uncertainties is difficult, if we assume a systematic uncertainty of  40\,\% due to these effects, then the reduced $\chi^2$ would be close to unity.  We adopt this uncertainty for all further calculations. }

{ In order to improve the fit quality, we incorporate} the above effects as a source of
systematic uncertainty 
in the $\gamma$-ray luminosity for each SFG in such a way that the reduced $\chi^2$ ($\chi^2/d.o.f.$) is of the order unity.
We { repeat}  the fit including this systematic uncertainty and
obtain best-fit  values for $\alpha$ and $\beta$ of: $\alpha = 1.27 \pm 0.06$ and $\beta = 39.47 \pm 0.08$. 
In Figure~\ref{fig:corrdetected} we show that the
correlation derived here is similar to the one derived in
\cite{lat_starforming}. Moreover, we also measure a
dispersion of 0.30 for the residuals of $L_{\gamma}$ about the best fit, 
in agreement with \cite{lat_starforming} and \cite{linden_2017}.

If we perform a fit to all the SFGs detected in our
  analysis, thus adding also the sources which have an offset with
  respect to the IR-$\gamma$-ray luminosity correlation, we find
  $\alpha = 1.23 \pm 0.06$ and $\beta = 39.55 \pm 0.07$. This result
  is compatible with the one derived using  the bona fide
  sample within uncertainties.





\begin{figure}
\centering
\includegraphics[width=0.5 \textwidth]{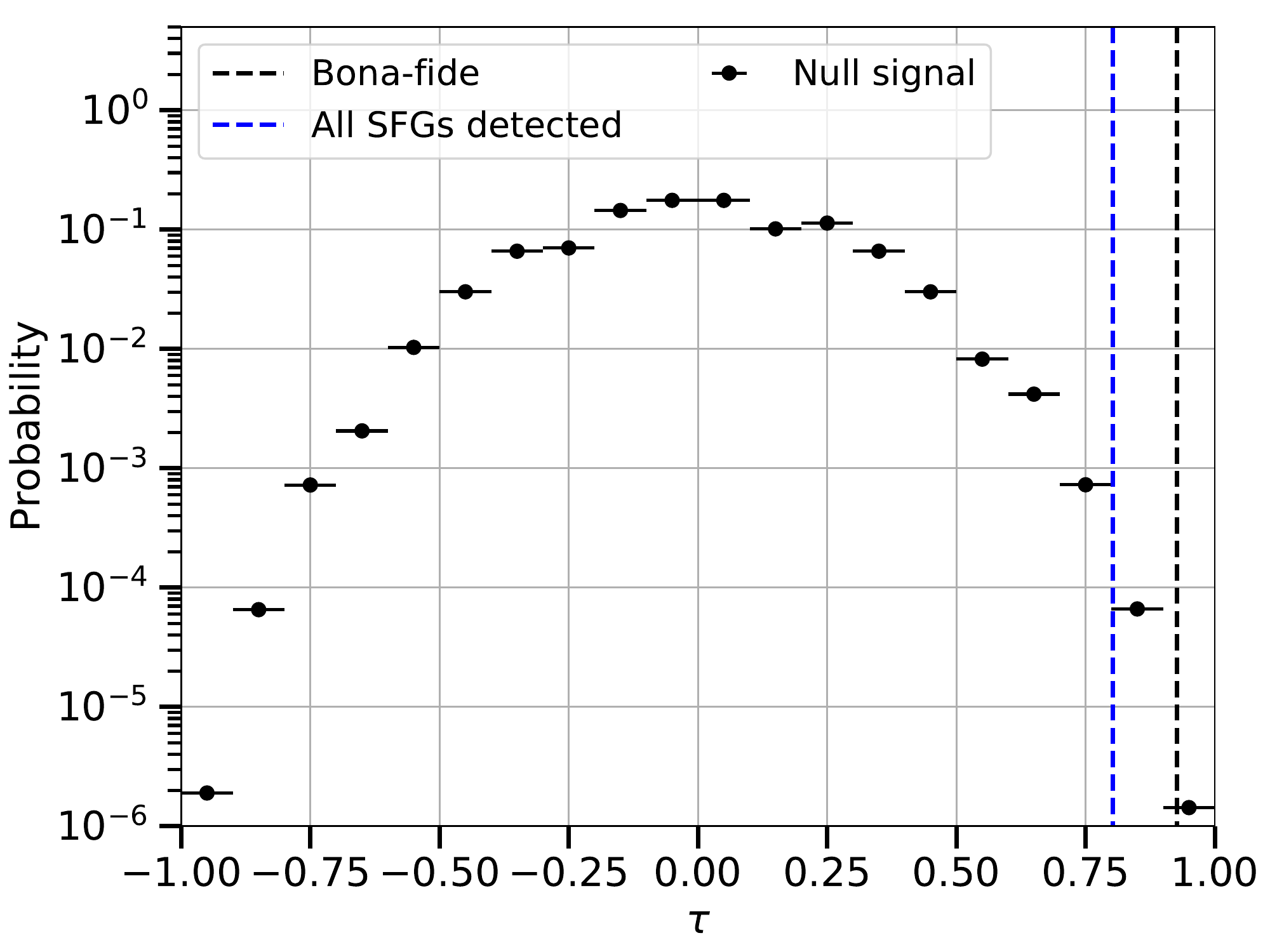}
\caption{Probability distribution for the $\tau$ parameter of
  the Kendall test performed with $30 \times 10^{6}$ realizations
  (black data). We display with a black (blue) dashed line the $\tau$ value found for the bona fide (complete) sample of detected sources.}
\label{fig:kendall}
\end{figure}

We test the significance of the correlation by performing a modified Kendall $\tau$ rank correlation test \citep{1996MNRAS.278..919A}. This is a {\it survival analysis}, which can be used for the analysis of partially censored data sets containing both detections and upper limits.
The parameter $\tau$ of the Kendall test varies between $[-1,+1]$ for negative and positive correlation.
We apply this test to our set of bona fide SFGs finding a value of
$\tau=0.92$. { When also including} NGC~2403 and NGC~3423
the value becomes $\tau=0.80$.
In order to convert this value into a significance we run Monte Carlo
realizations of samples of 11 and 13 SFGs with no correlations.
In particular we randomly draw $\gamma$-ray and IR luminosities in a range $\log_{10}{L_{\gamma}}\in[37,42]$ and $\log_{10}{L_{\rm{IR}}[L_{\odot}]}\in[7.5,12]$, respectively.
For each realization we calculate the Kendall $\tau$ parameter.
In Figure~\ref{fig:kendall} we show the histogram of the $\tau$ values
found for the $30 \times 10^{6}$ realizations considered. As expected this
is peaked at $\tau\approx 0$ since the simulations are based on the
absence of any correlation.
We use this histogram to find the probability distribution function
for the null hypothesis for the correlation and to find the p-value and significance associated to the $\tau$ values reported above.
The value of $\tau=0.92$ found  for the bona fide sample is associated to a p-value of $4.0\times 10^{-7}$ and a significance of 4.9\,$\sigma$.
On the other hand, the value of $\tau=0.80$ found with all
detected SFGs is associated to a p-value of $6.7\times 10^{-5}$ and a significance of 3.8\,$\sigma$.
As a comparison \cite{lat_starforming} found  a
p-value of 0.001.  Therefore, the significance of the correlation between $\gamma$-ray and
IR luminosities has grown significantly.

%
%
\subsubsection{Outliers}
\label{sec:outliers}

As reported above, NGC  2403 and NGC 3424
seem to be outliers in the $L_{\gamma}-L_{\rm IR}$ correlation.
These objects (particularly NGC 3424) are also somewhat above the
calorimetric limit\footnote{A galaxy becomes a CR calorimeter when CRs
  deposit most of their energy within the galaxy itself.} of
\cite{lat_starforming}, suggesting caution in interpreting the
observed signal as due entirely to star formation activity.
Indeed, these could be background fluctuations, background objects, or
sources whose $\gamma$-ray emission may be partially due to processes other than
star formation. The lack of variability and of strong radio emission
disfavors the hypothesis of a strong relativistic jet. Among the two
systems, only NGC 3424 is known to host a radio-quiet active galactic nucleus  \citep[AGN,][]{gavazzietal2011}; however the
$\gamma$-ray emission from radio-quiet AGN remains to be detected \citep{lat_seyfert_2012}.
{We also note that NGC~3424 and NGC~2403 were analyzed in
  \cite{peng2019} and \cite{xi2020}, where
  evidence of variability for both have been reported.}

As discussed in the previous section, the correlation
  between IR and $\gamma$-ray luminosities is not modified
  significantly when these two objects are included in the analysis. Indeed, the values of
  $\alpha$ and $\beta$ found in this case are compatible within
  1\,$\sigma$ with those derived using the bona fide sources (see Table~\ref{tab:correlation}).
Thus, while the origin of the $\gamma$-ray
  emission of these two objects may remain unclear, their inclusion in
  the analysis changes the results negligibly.

%
%
\subsection{L$_{\rm IR}$-L$_{\gamma}$ Correlation with the Stacking Analysis}
\label{sec:corr_stack}
Here we test whether the $\gamma$-ray emission of unresolved galaxies
is correlated with the IR emission in a way similar to what found in $\S$~\ref{sec:corr_det}
for the detected galaxies.
In order to do this, 
{ for every unresolved galaxy in our sample we generate}
a three-dimensional log-likelihood profile as a function of the
photon index ($\Gamma$) and the two parameters ($\alpha$ and $\beta$)
of the  L$_{\rm IR}$-L$_{\gamma}$ correlation. By subtracting the
log-likelihoods of the null hypotheses (i.e., there is no source) we transform these into $TS$
profiles. 

\begin{figure*}[ht!]
  \begin{center}
  \begin{tabular}{ll}
\hspace{-1cm}
  	 \includegraphics[scale=0.45,clip=true,trim=0 0 0
    0]{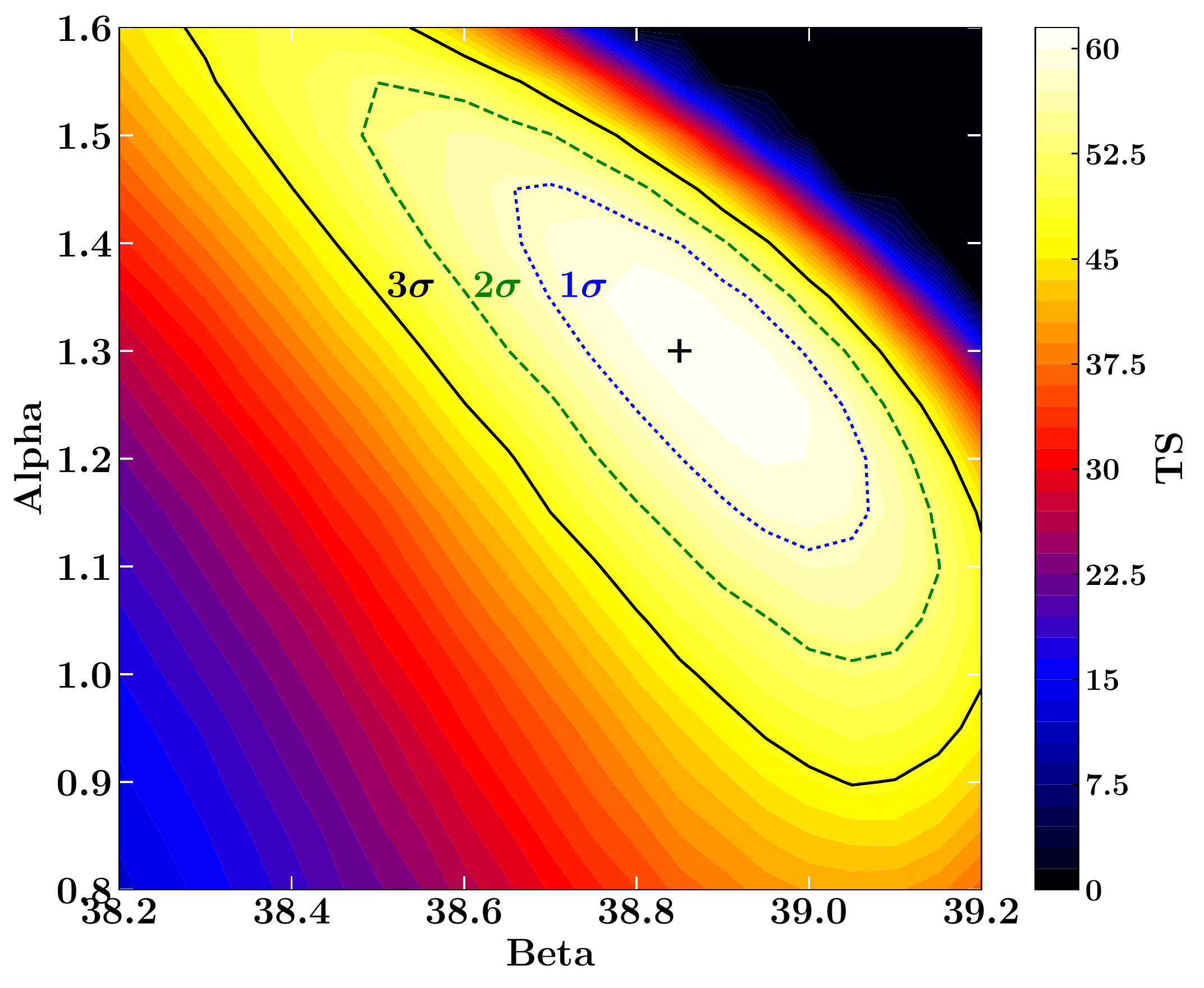} &
\hspace{-0.5cm}
	 \includegraphics[scale=0.45,clip=true,trim=0 0 0
    0]{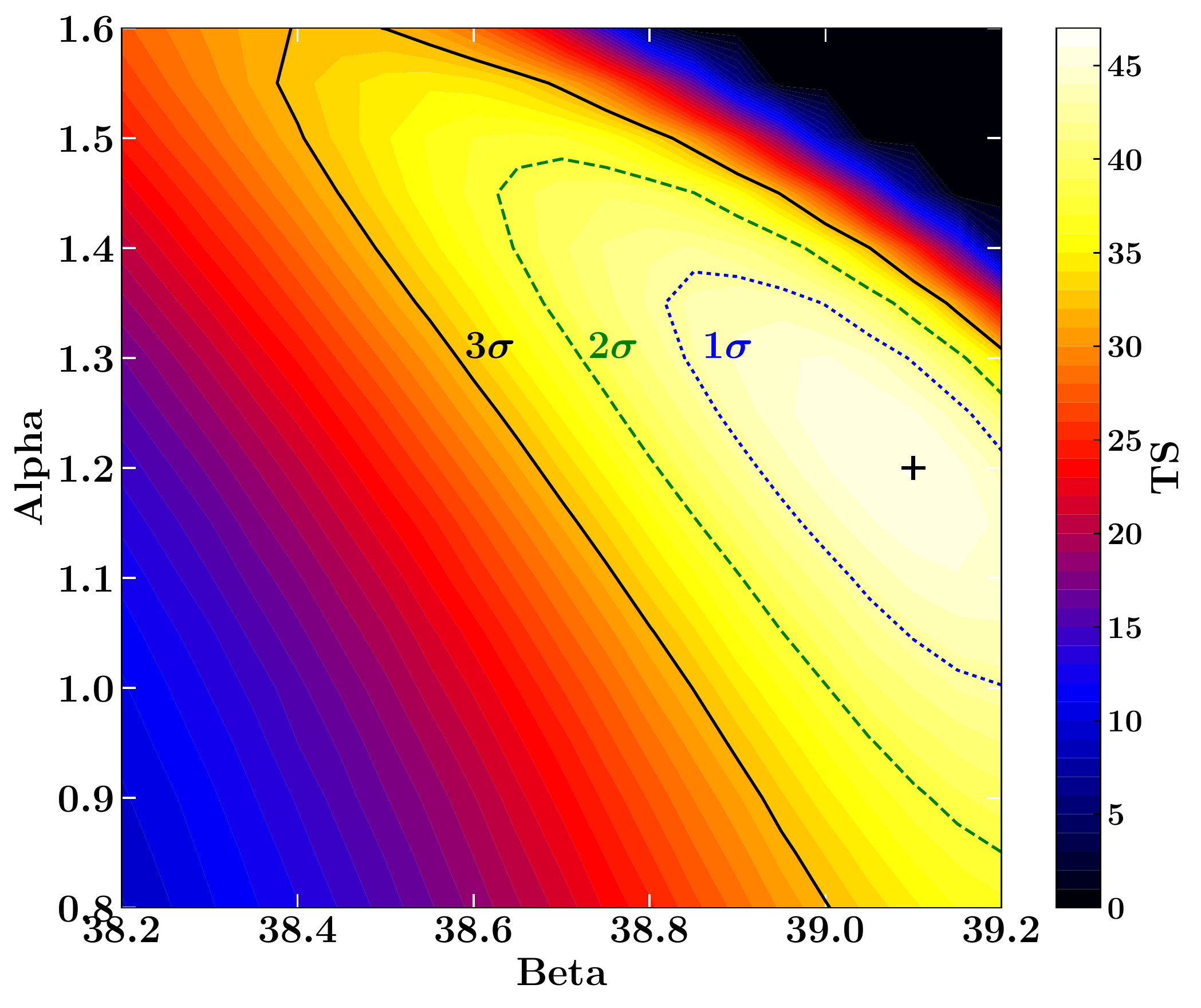} \\
\end{tabular}
  \end{center}
  \caption{Detection significance as a function of photon index and
    $\gamma$-ray flux  for all 574 undetected galaxies (left) and
     for the 56 undetected galaxies in the \cite{lat_starforming} sub-sample.
\label{fig:alpha_beta}}
\end{figure*}

The results for the full sample of 574 galaxies and the
ACK12 sub-sample are reported in Figure~\ref{fig:alpha_beta}. 
In both cases the $TS$ has grown substantially with respect
to the photon-index/flux stacking, indicating that the 3-parameter
fitting is better than the 2-parameter fitting. For the full sample and the ACK12
sample the $TS$ { values} are now $60$ (7.2\,$\sigma$) and $45$ (6.12\,$\sigma$), respectively. 
The best-fit parameters, $\alpha$ and $\beta$, are in very good
agreement with those derived for detected galaxies (see Table~\ref{tab:correlation}). We thus
conclude that the $\gamma$-ray luminosity of unresolved SFGs
correlates with their IR luminosity.

\begin{deluxetable}{lccc}
 \tablewidth{0pt}
 \tablecaption{Best-fit Parameters of the L$_{\rm IR}$-L$_{\gamma}$ Correlation
\label{tab:correlation}}
\tablehead{\colhead{Sample} & \colhead{\# Galaxies} &\colhead{$\alpha$} &
  \colhead{$\beta$} 
 }
 \startdata
Detected  (bona fide)        &  11 &   $1.27^{+0.06}_{-0.06}$ &
$39.47^{+0.08}_{-0.08}$  \\
Detected (all) & 13 &  $1.23^{+0.06}_{-0.06}$ &
$39.55^{+0.07}_{-0.07}$  \\
Unresolved & 574 & 1.30$^{+0.10}_{-0.10}$ &  38.85$^{+0.15}_{-0.11}$\\
Unresolved (ACK12)             & 56   & 1.20$^{+0.10}_{-0.15}$ &
39.10$^{+0.10}_{-0.10}$\\
\hline
Detected + Unresolved & 587   & 1.15$^{+0.08}_{-0.03}$ &
39.20$^{+0.06}_{-0.05}$\\
 \enddata
 \end{deluxetable}

\subsection{Combined Constraints}

The constraints on the $\alpha$ and $\beta$ parameters of Eq.~\ref{eq:corr}
from resolved and unresolved sources can be combined to obtain a
correlation that describes both populations. This is done by
transforming the $\chi^2$ of the fit { described in
  $\S$~\ref{sec:corr_det} for}
resolved objects into a log-likelihood as $\frac{1}{2}\chi^2 = 
-logL$. This can be summed with the log-likelihood profile derived for
unresolved galaxies, yielding best-fit parameters of
$\alpha=1.15 ^{+0.08}_{-0.03}$ and $\beta=$39.20$^{+0.06}_{-0.05}$.

%
%
\section{The $\gamma$-ray  Spectral Energy Distribution of Galaxies}
\label{sec:spec}

In this section we aim to understand whether SFGs
have a common SED { in the $\gamma$-ray band}.

\subsection{Combined SED of individually detected SFGs}
\label{sec:sed_detected}
One of the final goals of this work is { to determine the
  contribution from} SFGs to the EGB and to the astrophysical neutrino flux measured by IceCube.
One key ingredient for this estimate is the SED of SFGs. Therefore, we consider the detected SFGs and perform a combined likelihood fit to their SEDs.

We consider a smoothly broken power-law (SBPL) model given by:
 \begin{equation}
 \frac{dN}{dE} = K \left( \frac{E}{E_0} \right)^{-\Gamma_1} \left (  1 +\left ( \frac{E}{E_0} \right)^{-\Gamma_1+\Gamma_2} \right)^{-\beta},
 \label{eq:LP2}
 \end{equation}
where $E_0$ is left as a free parameter. In this analysis, the
normalization $K$ is free to vary in the fitting for the galaxies’ SED, but all other
parameters are common to all objects.
The combined analysis is done by considering the SED results in the
form of likelihood profiles as a function of $dN/dE$ in each energy bin
and combining them for all the sources of the bona fide sample except
the SMC and the LMC. These two galaxies are removed from this analysis
because they have the lowest IR luminosity, but the highest $TS$ and
would drive the result of the likelihood analysis.
In the fit we also include, to best characterize the high-energy SED
of SFGs, the TeV data points of NGC~253 detected by
H.E.S.S. \citep{2005A&A...442..177A} and M82 detected by VERITAS
\citep{2009Natur.462..770V}.

\begin{figure*}[ht!]
  \begin{center}
  \begin{tabular}{ll}
\hspace{-1cm}
\includegraphics[width=0.5 \textwidth]{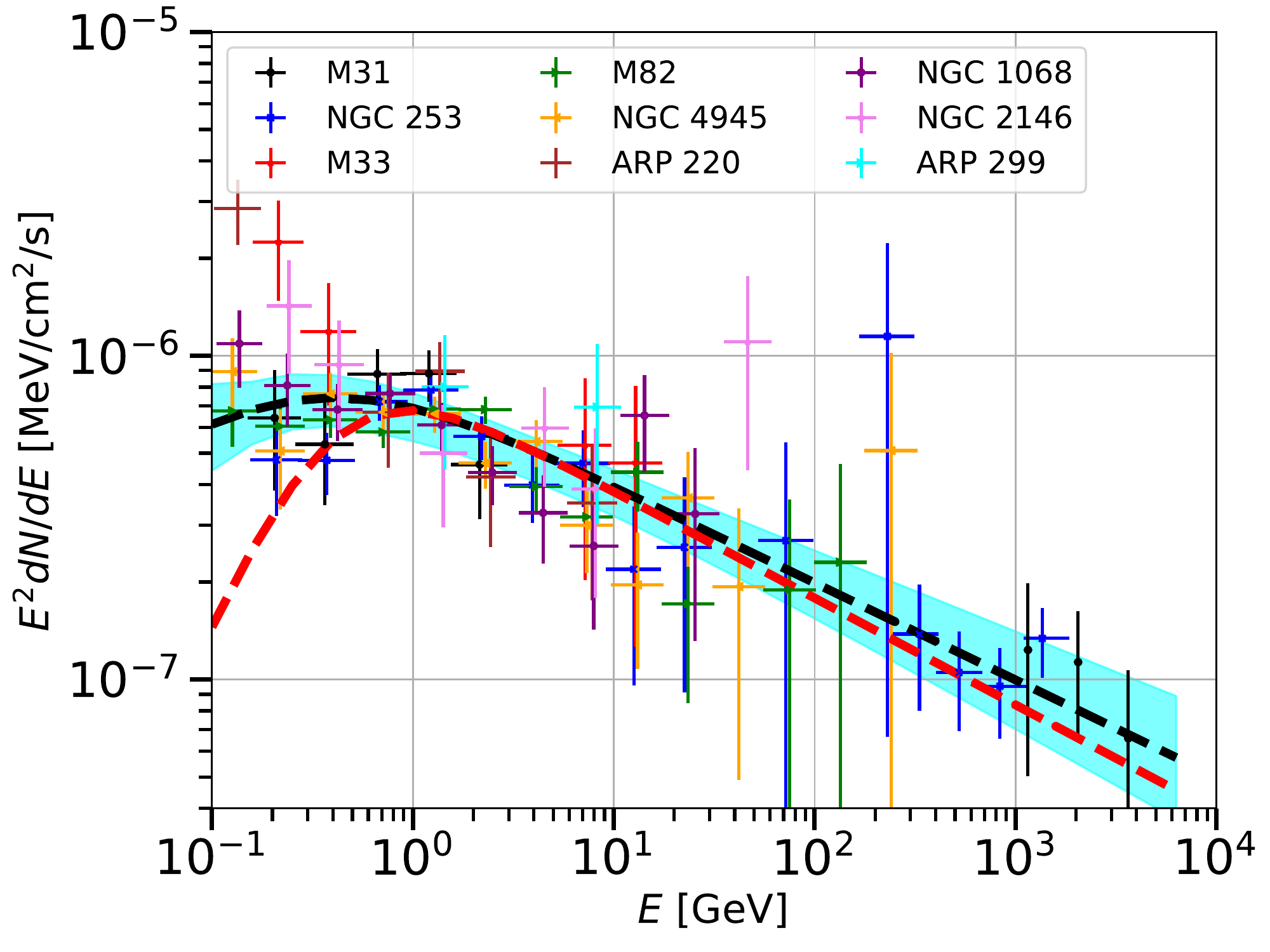}
  	 \includegraphics[scale=0.5,clip=true,trim=0 0 0
    0]{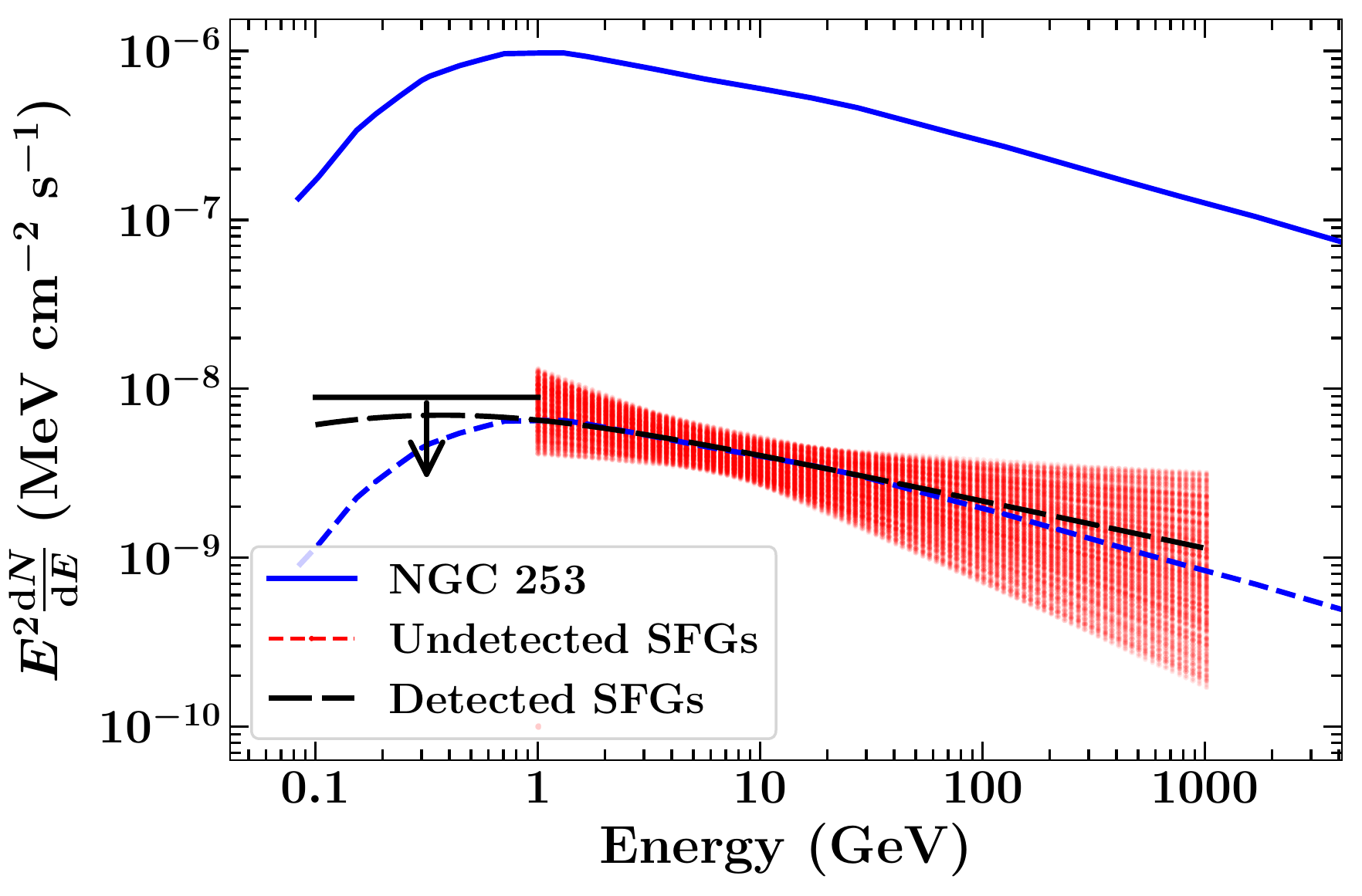} 
\end{tabular}
  \end{center}
  \caption{{\bf Left Panel}: Composite SED model of resolved galaxies (black
    dashed line) with its 1\,$\sigma$ uncertainty band (cyan
    region). For clarity, the SED
    of each galaxy has scaled to a common value at t 1\,GeV and upper limits are
    not shown. The red-dashed line is the physical model fit to the 
LAT and H.E.S.S. data of NGC~253 \citep{hess_ngc253_2018}.
 {\bf Right Panel}:
SED of unresolved galaxies (red hatched region and black {95\,\%
C.L.} upper  limit) as determined from the stacking
    analysis of $\S$~\ref{sec:sed_stack} compared to the hadronic emission
    model (blue line) that reproduces the LAT and H.E.S.S. data of NGC~253
    \citep{hess_ngc253_2018} rescaled (blue dashed line) to the SED of
    unresolved galaxies. The black dashed line shows the best-fit SED model
    of detected SFGs (see left panel). \label{fig:sedcombined}}
\end{figure*}

The best-fit values for the SED parameters are: $\Gamma_1=2.28\pm0.07$,
$\Gamma_2=1.60\pm0.2$, $\log_{10}{E_0 [\rm{GeV}]}=-0.50\pm0.15$ and $\beta=0.75\pm0.20$. 
The values of the SED parameters do not change significantly if we exclude  M31 and M33.
The left panel of figure~\ref{fig:sedcombined} shows the composite SED model (with its uncertainty)
together with the SEDs of all { detected} galaxies scaled to a
common value at 1\,GeV. We find
that other SED shapes, like PL or LP, are disfavored over the SBPL at
the 4.6\,$\sigma$  and 3.1\,$\sigma$ level, respectively.

%
%

\subsection{Combined SED of Unresolved Galaxies}
\label{sec:sed_stack}
The results of $\S$~\ref{sec:flux_stack} show that {above}
$>$100\,MeV the
spectrum of unresolved galaxies is hard, displaying a power-law index
of $\Gamma\approx2.1-2.3$. However, the true SED of SFGs may be
more complex than a single power law and { a} hard power-law  index may
artificially originate when fitting around the SED peak. Moreover,  a
hard $\gamma$-ray spectrum could have interesting implications for the EGB and the
neutrino background.

\begin{figure*}[ht!]
  \begin{center}
  \begin{tabular}{ll}
\hspace{-1cm}
  	 \includegraphics[scale=0.45,clip=true,trim=0 0 0
    0]{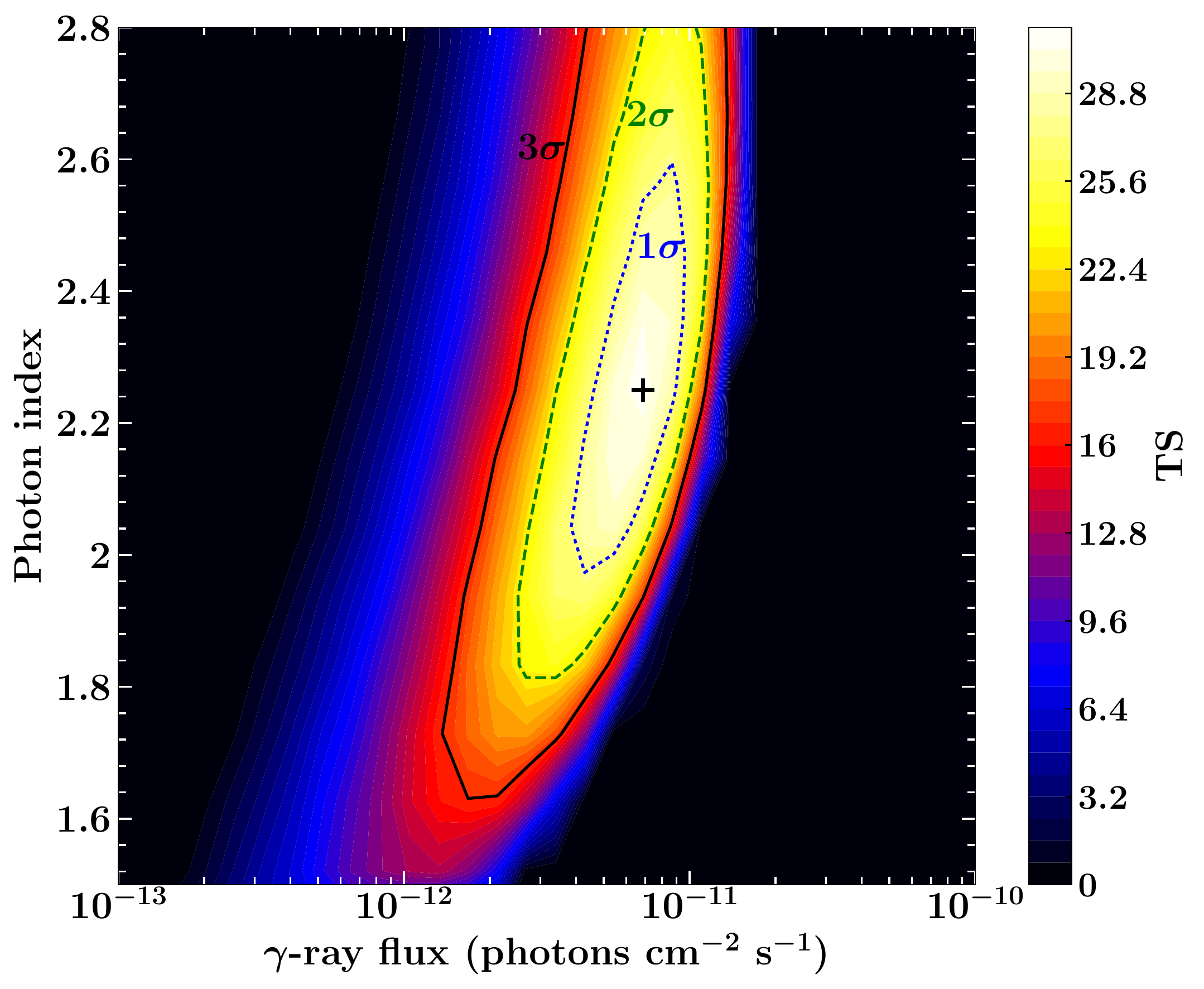} &
\hspace{-0.5cm}
	 \includegraphics[scale=0.45,clip=true,trim=0 0 0
    0]{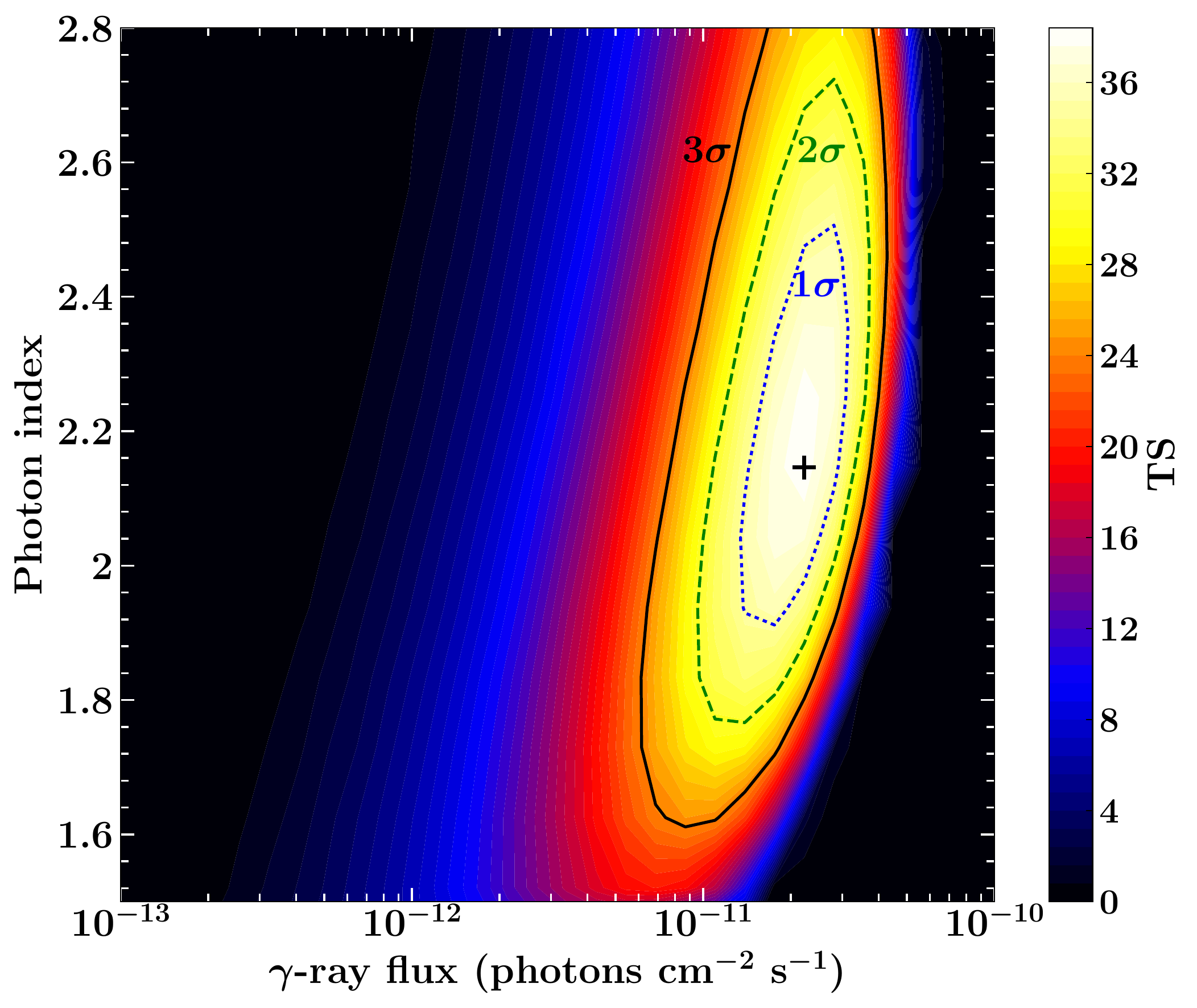} \\
\end{tabular}
  \end{center}
  \caption{Detection significance as a function of photon index and
    $\gamma$-ray flux ({ for energies above} 1 GeV)  for all 574 undetected galaxies (left) and
    for the 56 undetected galaxies in the \cite{lat_starforming}
    subsample (right).
\label{fig:1gev_stack}}
\end{figure*}

In order to test the shape of the SED of unresolved galaxies, we repeat the stacking
analysis in two separate energy ranges: 0.1--1.0\,GeV
and 1.0--800\,GeV  and display the results in
Figure~\ref{fig:1gev_stack} .
The lower energy range yields a 
$TS$$\approx$0, which is translated into a 95\,\% confidence level
upper limit (see Figure~\ref{fig:1gev_stack}).
The energy range above
$1$\,GeV confirms the results shown in
$\S$~\ref{sec:flux_stack}. {In particular, the best-fit photon
  index is 2.26$^{+0.28}_{-0.16}$.}
The
$TS$ is larger because the less-intense background and smaller PSF at
these energies contribute to increase the signal-to-noise ratio. The
analysis presented in this section shows
that the average SED of SFGs does not break above 1\,GeV and
remains compatible with a power law with $\Gamma\approx2.3$.
This is compatible with the SED found in Sec.~\ref{sec:sed_detected} for the detected galaxies.

The right panel of Figure~\ref{fig:sedcombined} shows the SED of unresolved galaxies, derived from
the stacking analysis, compared to
{ that} of NGC~253 \citep{hess_ngc253_2018} and the composite SED model
of $\S$~\ref{sec:sed_detected}. Both models are good representations of
the SED of unresolved SFGs and show that above 1\,GeV the SED of
galaxies (both resolved and unresolved) is compatible with a power law
with a photon index $\Gamma\sim2.3$.  { However,  Figure~\ref{fig:sedcombined}
  shows that our fit provides a
better representation for the SED of detected galaxies than the model
of \cite{hess_ngc253_2018} and as such is adopted here.}
{ In the next section we use} 
the composite SED model of
detected SFGs as the representative shape for the SED of all SFGs, in order to predict their contribution to the EGB
and IceCube astrophysical neutrino flux \citep{PhysRevLett.113.101101}.

%
%
\section{Implications for the EGB and IceCube neutrino background}
\label{sec:EGBnu}

The EGB certainly { contains} the emission of SFGs.
In the previous sections, we have reported the detection of  a 11
SFGs and the emission from unresolved galaxies.
Here we calculate the contribution of this source population to the {\it Fermi}-LAT EGB \citep{lat_egb2} and to the IceCube astrophysical neutrino flux \citep{PhysRevLett.113.101101}.

The diffuse $\gamma$-ray flux due to the whole population of SFGs can be evaluated as follows:
    \begin{eqnarray} 
    \label{egrbdef}
\frac{d^2F(\epsilon)}{d\epsilon d\Omega} &=& 
	  \int^{z_{max}}_0 \frac{d^2V}{dz d\Omega} dz \int^{L_{\gamma,max}}_{L_{\gamma,min}} \frac{d F_{\gamma}}{d\epsilon}  \frac{dL_{\gamma}}{L_{\gamma} \ln(10)} \cdot \\ 
    &\cdot& \rho_{\gamma}(L_{\gamma},z) \exp{(-\tau_{\gamma,\gamma}(\epsilon,z))}.  \nonumber
    \end{eqnarray}
The minimum $\gamma$-ray luminosity value is set to $10^{35}$ erg
s$^{-1}$, the maximum is set to $10^{45}$ erg s$^{-1}$ and {the maximum
redshift is set to $z_{max}=5$}. These
limits are wide enough to contain the emission of the entire galaxy population.
The quantity $d F_{\gamma}/d\epsilon$ is the intrinsic photon flux at
energy $\epsilon$, for a source with a $\gamma$-ray luminosity $L_{\gamma}$ and redshift $z$, and $\rho_{\gamma}$ is the $\gamma$-ray luminosity function.
High-energy $\gamma$ rays ($\epsilon > 20$ GeV) propagating in the
Universe can be absorbed in interaction with the extragalactic
background light (EBL) photons
\citep[e.g.,][]{1966PhRvL..16..252G,stecker92,finke2010},
and    $\tau_{\gamma,\gamma} (\epsilon, z)$ is the 
optical depth  for such process.
In this study we adopt the attenuation model of \cite{finke2010},
which was found to be in good agreement with recent measurements of the EBL
optical depth \citep{lat_ebl2018}. 
The $\gamma$-ray absorption creates electron-positron pairs, 
which can inverse-Compton scatter off the cosmic microwave background photons, yielding 
secondary cascade emission at lower  $\gamma$-ray energies. The
cascade flux is, however, negligible because the SED of SFGs at high
energy has a photon index greater than 2; for this reason we decide to neglect this
contribution. We also do not consider the internal (to the source)
absorption caused by a galaxy's interstellar radiation fields
because { it is}  important { only}  above 10\,TeV \citep{inoue2011}.

The $\gamma$-ray luminosity function is related to the IR one using:
\begin{eqnarray}
     \label{rho_gamma1}    
        \rho_{\gamma}(L_{\gamma}) = k  \rho_{{\rm IR}}\left(L_{{\rm IR}}(L_{\gamma})\right) \frac{d\log L_{{\rm IR}}}{d\log L_{\gamma}},
\end{eqnarray}
where $k$ is assumed to be equal to 1 and the factor $ \frac{d\log
  L_{{\rm IR}}}{d\log L_{\gamma}}$ { is} computed from the
correlation between the $\gamma$-ray and IR luminosities reported in Sec.~\ref{sec:corr_stack} for the combined analysis of detected and undetected sources in our sample.
We consider the IR luminosity function provided by {\it Spitzer} observations of the VIMOS VLT Deep Survey and GOODS fields \citep{rodighiero10}.
We also use the total IR luminosity function derived from the
observations of the {\it Herschel} GTO PACS Evolutionary Probe Survey,
in combination with the HERschel Multi-tiered Extragalactic Survey
(HerMES) data \citep{Gruppioni:2013jna}. 
\cite{Gruppioni:2013jna} provide the IR luminosity functions for normal spiral galaxies, starburst galaxies, and galaxies which contain an either obscured or low-luminosity AGN (SF-AGN). Finally they also report the IR luminosity function for galaxies containing a powerful starburst component, mainly responsible for their far-IR emission, and an AGN component that contributes significantly to the mid-IR (AGN1 and AGN2). In our calculation with the \cite{Gruppioni:2013jna} IR luminosity function we include the spiral, starburst, and SF-AGN components.

Since $\gamma$ rays and neutrinos are produced together in
star-forming galaxies via the production and decay of energetic pions
from hadronic CR interactions, the flux of $\gamma$ rays and
$\nu$ (per flavor) are related as:
\begin{eqnarray}
     \label{rho_gamma2}    
       \Phi_{\nu}(E_{\nu}) =   \frac{K_{\pi}}{4} \Phi_{\gamma} (E_{\gamma}),
\end{eqnarray}
where $K_{\pi}$ is the relative charged-to-neutral pion rate. We assume for proton-gas ($pp$) collisions $K_{\pi}=2$ which corresponds to an equal contribution of $\pi^0$ and $\pi^{\pm}$. The average energies of $\gamma$ rays and neutrinos are related as $E_{\gamma} \approx 2E_{\nu}$.

\begin{figure*}
\centering
\includegraphics[width=0.70 \textwidth]{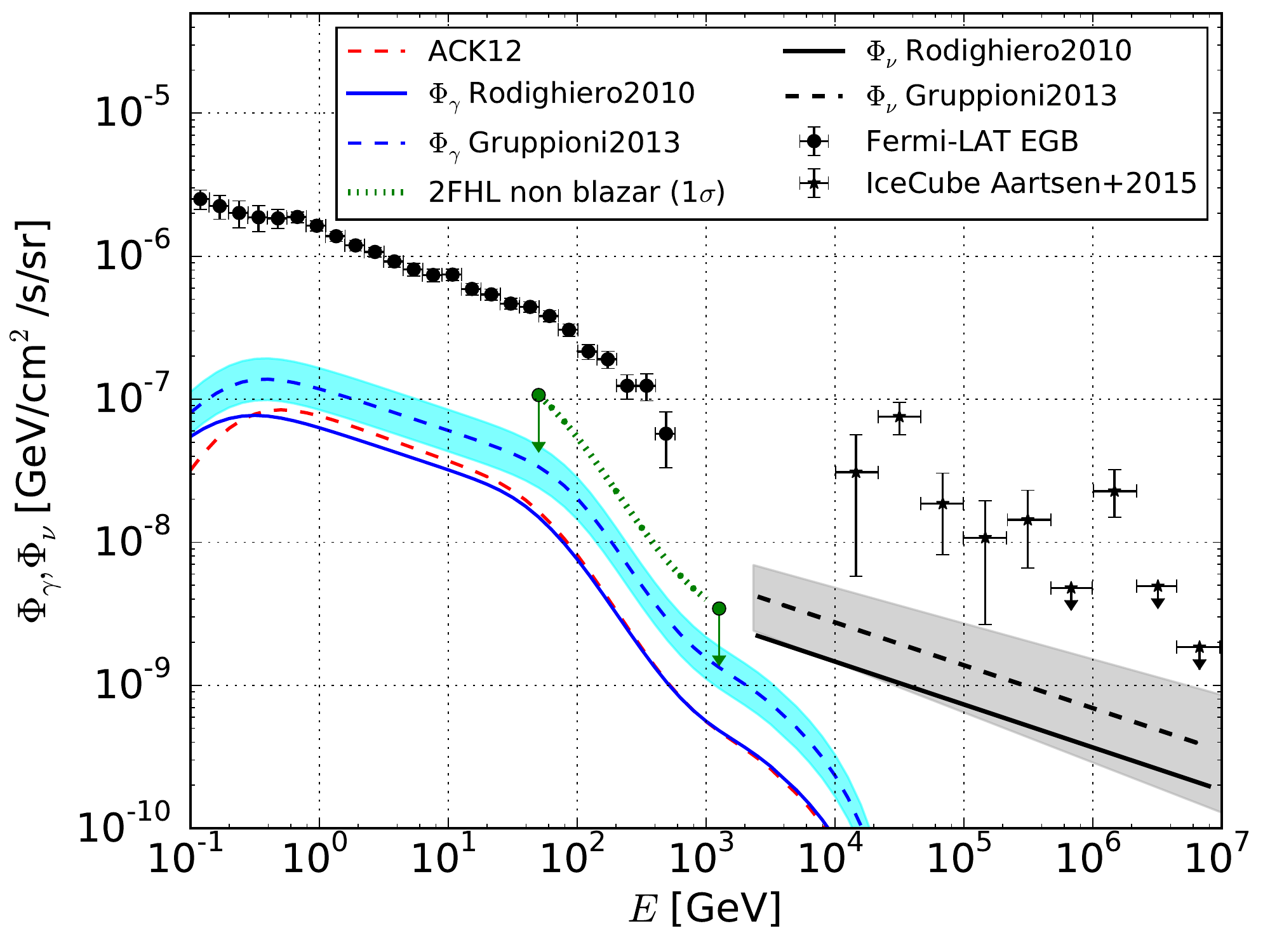}
\caption{Gamma-ray intensity (blue solid line and cyan 1\,$\sigma$ band) and $\nu$
  (black solid line and gray 1\,$\sigma$ band) intensity from SFGs compared to the {\it
    Fermi}-LAT EGB \citep{lat_egb2} and astrophysical neutrino flux measured by
  IceCube \citep{PhysRevLett.113.101101}. We also show the estimate for the SFG contribution to the
  EGB as derived in \cite{lat_starforming} (dashed red line). The
  dashed green line is the 1\,$\sigma$ upper limit for the non-blazar
  contribution to the EGB above 50\,GeV \citep{ackermann16_egbres}.} 
\label{fig:gammanuflux}
\end{figure*}

In Figure~\ref{fig:gammanuflux} we show the resulting prediction for the
contribution of SFGs to the EGB and astrophysical $\nu$ flux.
SFGs on average contribute to the {\it Fermi}-LAT EGB at the level of $5\%$
and to the IceCube $\nu$ flux at the level of $3\%$ \citep[see also][]{bechtol2017,sudoh2018}.
Our results for both the IR luminosity functions are compatible with the $1\sigma$ upper limits for a non-blazar contribution to the EGB above 50 GeV derived in \cite{ackermann16_egbres}.
Finally, we note that the derived luminosity function under-predicts
the current
number of detected galaxies at $\gamma$-ray energies. This is due to the local
overdensity, as most of our galaxies are detected within
50\,Mpc. However, this does not affect our results, since 
 the dominant contribution to the background flux  comes
from more distant galaxies. 

%
%
\section{Conclusions}
\label{sec:conclusions}

We have used almost 10 years of {\it Fermi}-LAT data to
characterize the $\gamma$-ray emission of SFGs and
calculate their contribution to the {\it Fermi}-LAT EGB and IceCube
$\nu$ flux. Our findings are summarized as follows:
\begin{itemize}
\item We have reported the detection of 11 bona fide $\gamma$-ray
  emitting galaxies and 2 galaxy candidates, a sizable improvement over the 7
  systems (not counting the Milky Way) reported in \cite{lat_starforming}.
\item Similarly to \cite{linden_2017}, we have shown that the emission
  of unresolved galaxies is detected by {\it Fermi}-LAT with a
  significance in the $4.4-5.1$\,$\sigma$ range (assuming these
  objects can be characterized by an average flux and spectral index). This significance
  increases to 7.2\,$\sigma$ when unresolved galaxies are allowed to
  follow the $L_{\gamma}-L_{\rm IR}$ correlation (see next bullet point).
\item The $\gamma$-ray luminosity of both detected and unresolved galaxies
{ has been} found to correlate with the total IR luminosity, as in previous
works \citep{lat_starforming,linden_2017}. The significance of this
correlation (using only the bona fide galaxies) is now near 5\,$\sigma$.
\item Using both detected and unresolved objects, we { have} constrained the
  average SED of $\gamma$-ray emitting galaxies, which was found
  (above 1\,GeV) to be compatible with
  a power law with a photon index of $\approx2.2-2.3$.

\item Finally, we estimated that the contribution of star-forming galaxies to {\it Fermi}-LAT EGB and IceCube $\nu$ flux  is 5\% and 3\%, respectively.
This scenario is compatible with the current estimate for blazar and radio galaxy contributions to the EGB \citep[see e.g.][]{DiMauro:2013zfa,ajello15,ackermann16_egbres,dimauro18}.

\end{itemize}

\clearpage
\acknowledgments
MA acknowledges funding under NASA contract 80NSSC18K1718.

The \textit{Fermi} LAT Collaboration acknowledges generous ongoing support
from a number of agencies and institutes that have supported both the
development and the operation of the LAT as well as scientific data analysis.
These include the National Aeronautics and Space Administration and the
Department of Energy in the United States, the Commissariat \`a l'Energie Atomique
and the Centre National de la Recherche Scientifique / Institut National de Physique Nucl\'eaire et de Physique des Particules in France, the Agenzia 
Spaziale Italiana and the Istituto Nazionale di Fisica Nucleare in Italy, 
the Ministry of Education, Culture, Sports, Science and Technology (MEXT), 
High Energy Accelerator Research Organization (KEK) and Japan Aerospace 
Exploration Agency (JAXA) in Japan, and the K.~A.~Wallenberg Foundation, 
the Swedish Research Council and the Swedish National Space Board in Sweden.
Additional support for science analysis during the operations phase 
is gratefully acknowledged from the Istituto Nazionale di Astrofisica in 
Italy and the Centre National d'\'Etudes Spatiales in France.
 This work performed in part under DOE Contract DE-AC02-76SF00515.

\facility{Fermi/LAT}

%
%
\appendix

%
%
\section{Details For Individual Galaxies}
\label{sec:appA}

\subsubsection{Previously detected SFGs}
The brightest and closest SFGs detected by the LAT are the Large and Small Magellanic Clouds.
They are detected as extended sources in {\it Fermi}-LAT data with extensions of about $3.0^{\circ}$ and $1.5^{\circ}$, respectively.
We consider for these two sources the spatial templates as given in the Preliminary 8 years list (FL8Y) \citep{TheFermi-LAT:2015lxa}.

Other bright SFGs detected in $\gamma$ rays include M82 and NGC 253 (which are both at distance of 3.2 Mpc), NGC 4945 (at a distance of 4.0 Mpc), and NGC 1068 (at a distance of 10.6 Mpc). 
M82 is the largest galaxy of the M81 group in the Ursa Major
constellation and it is the closest galaxy that hosts a starburst
nucleus. Its star formation rate is about 10\,M$_{\odot}$ yr$^{-1}$
and it has been detected at TeV energies by VERITAS  \citep{2009Natur.462..770V}.
NGC 253 is a giant barred spiral galaxy with a central starburst region.
{\it NuSTAR} and {\it Chandra} observations pointed out the presence
of X-ray sources but there
is no clear evidence indicating that these sources are associated with a
a low-luminosity AGN \citep{2013ApJ...771..134L}.
The star formation rate of NGC 253 has been estimated to be about 5M\,$_{\odot}$􏰃yr$􏰈^{-1}$ \citep{2002ApJ...574..709M}.
NGC 253 has also been detected  by H.E.S.S. at TeV energies with an
SED given by a power law with index of 2.14 \citep{2012ApJ...757..158A}. 
NGC~4945 is a circumnuclear starburst galaxy that hosts an obscured
Compton-thick Seyfert 2 nucleus that is among the brightest in the hard
X-ray range \citep{2012MNRAS.423.3360Y}.
In \cite{Wojaczynski:2017muc} the hint of a correlation between the
$\gamma$-ray and X-ray emission has been reported, suggesting that the
$\gamma$-ray emission may be associated with the Seyfert component.
The lightcurve of this source (see $\S$~\ref{sec:lc} and Fig.~\ref{fig:lc1}) 
is particularly stable with no indications of variability. 
The apparent lack of $\gamma$-ray variability favors the interpretation of emission dominated by starburst processes.
NGC~1068 is a source that exhibits both starburst and AGN activity,
whose Seyfert 1 nucleus has been discovered in polarized light \citep{1985ApJ...297..621A}.
Models that explain the $\gamma$-ray emission from this source  also
take into account  the emission from the active nucleus
\cite[see, e.g., ][]{lenain2010,2014ApJ...780..137Y,eichmann2016}.

Among the faintest SFGs are M31, NGC 2146, and Arp~220 located at around 0.8 Mpc, 19.6 Mpc and 79.9 Mpc, respectively.
M31 is the nearest massive spiral galaxy and it
has a star formation rate of 0.35--1.0\,M$_{\odot}$ yr$^{-1}$ \citep{2009A&A...505..497Y}.
This source is found to be extended in our $\gamma$-ray analysis with
a $TS$ of extension of 21 and an angular size of $0.45 \pm 0.10$
deg. The extension is modeled using a   disk and yields a
consistent morphology to what found in \cite{Ackermann:2017nya}.
NGC 2146 is a starburst galaxy located at about 19.6 Mpc from Earth,
whose starbursting episode is 
believed to have been started approximately 800 Myr ago when two
galaxies collided  (see \cite{refId0} and references therein). 
NGC 2146 shows no evidence of AGN activity in optical
\citep{1997ApJS..112..315H} or in  mid-IR spectra \citep{2009ApJS..184..230B}.
This source exhibits a relatively high SFR of 7.9 M$_{\odot}$\,yr$^{-1}$ \citep{2011PASP..123.1347K}.
Arp~220 is the farthest SFG detected by the LAT and it is a merger of
two galaxies containing two dense nuclei separated by $\sim$350\,pc
(see, e.g., \cite{1998ApJ...492L.107S}). 
Both nuclei have high SFRs and dense molecular gas. 
This system has a total SFR of 240$\pm$ 3.0\,M$_{\odot}$yr$^{-1}$,
calculated from the far-IR luminosity \citep{2003MNRAS.343..585F}.

\subsubsection{Newly detected SFGs}
The newly detected SFGs are: M33 \citep[see
also][]{karwin19,dimauro19}, NGC 2403, UGC 11041, NGC 3424, and Arp
299.
M33 is located roughly 870\,kpc from Earth. It is a satellite galaxy
of M31 and is the first extragalactic satellite to be detected in
$\gamma$ rays.
M33 has a SFR in the range 0.26-0.70 \,M$_{\odot}$ yr$^{-1}$ \citep{2007A&A...473...91G}.

NGC 2403 is located at about 3.4 Mpc from Earth. It is an H\,{\sc i}-dominated, low-mass spiral galaxy, and its HI gas disc extends beyond its optical disc \citep{2017MNRAS.469.1636K}. 
The total SFR has been measured to be about 1.3
M$_{\odot}$ yr$^{-1}$ \citep{2003PASP..115..928K}.
Our analysis shows that the $\gamma$-ray peak is 0.048$^{\circ}$ away
from the nominal center of the galaxy, with  the 68\% positional error of the
source being 0.036$^{\circ}$.

NGC 3424 is located at about 29\,Mpc from Earth. It is considered
to be a SFG  \citep{2016ApJ...817...76M} with a SFR
of about 4.5\,M$_{\odot}$ yr$^{-1}$.
In \cite{gavazzietal2011} the optical spectrum
shows the presence of an AGN in the center of the Galaxy.

Arp 299 is located at a distance of 47.7\,Mpc from Earth and it is an interacting system composed of two individual galaxies (IC 694 and NGC 36907) in an early dynamical stage.
Powerful starburst regions with SFRs of about 100 \,M$_{\odot}$
yr$^{-1}$ have been identified in the system \citep{2000ApJ...532..845A}.
Optical spectroscopic studies have classified IC 694 as starburst
galaxy and NGC 3690 as starburst/LINER galaxy \citep{1998ApJS..119..239C}, and mid-IR observations classified the system as starburst \citep{2000A&A...359..887L}.
However, there are hints for the presence of AGN activity \citep{2002ApJ...581L...9D,2004ApJ...600..634B,2004A&A...414..845G}.

\subsubsection{Lightcurves of detected sources}
\label{sec:lc}
\begin{figure}
\centering
\includegraphics[width=0.48 \textwidth]{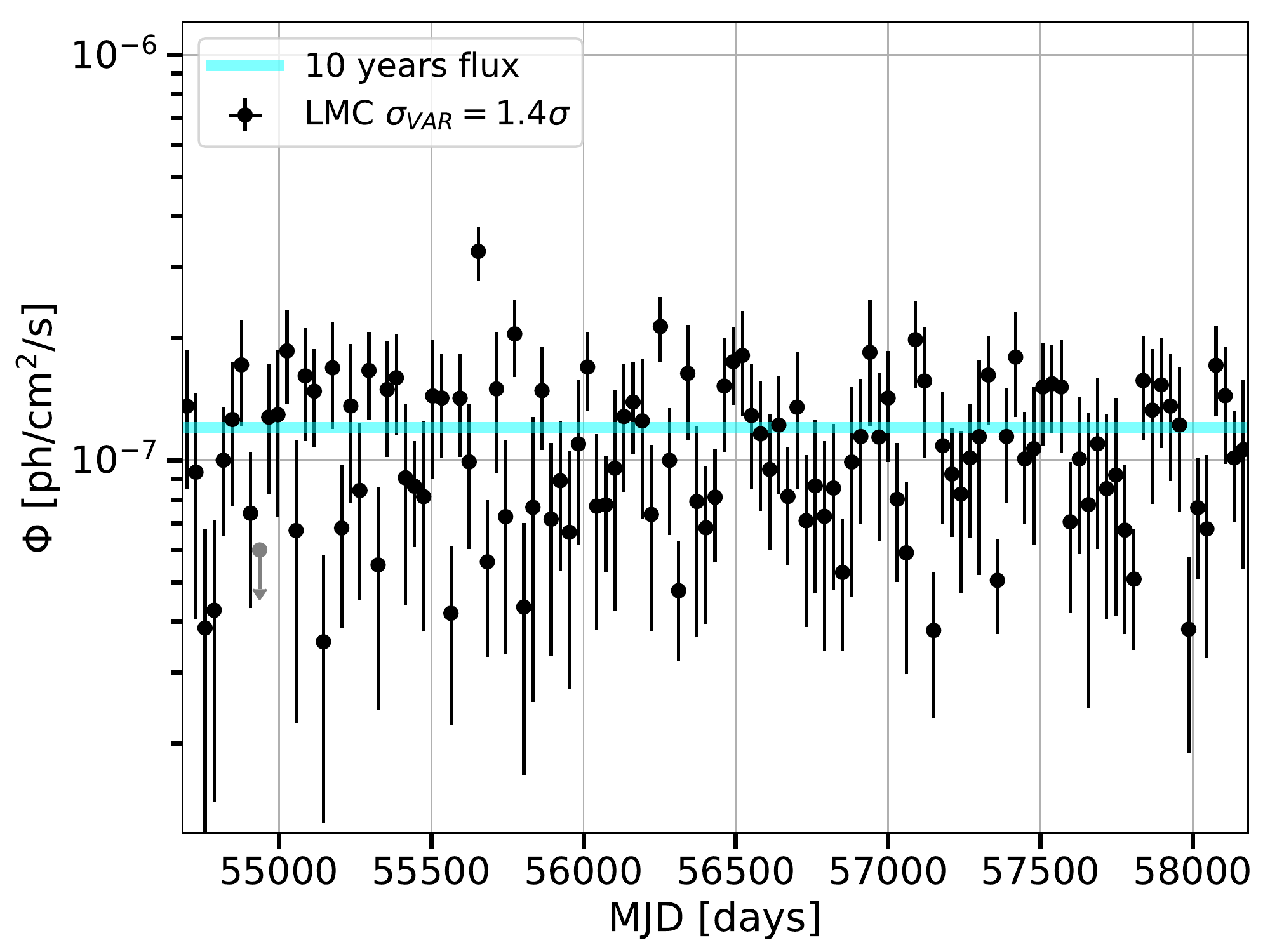}
\includegraphics[width=0.48 \textwidth]{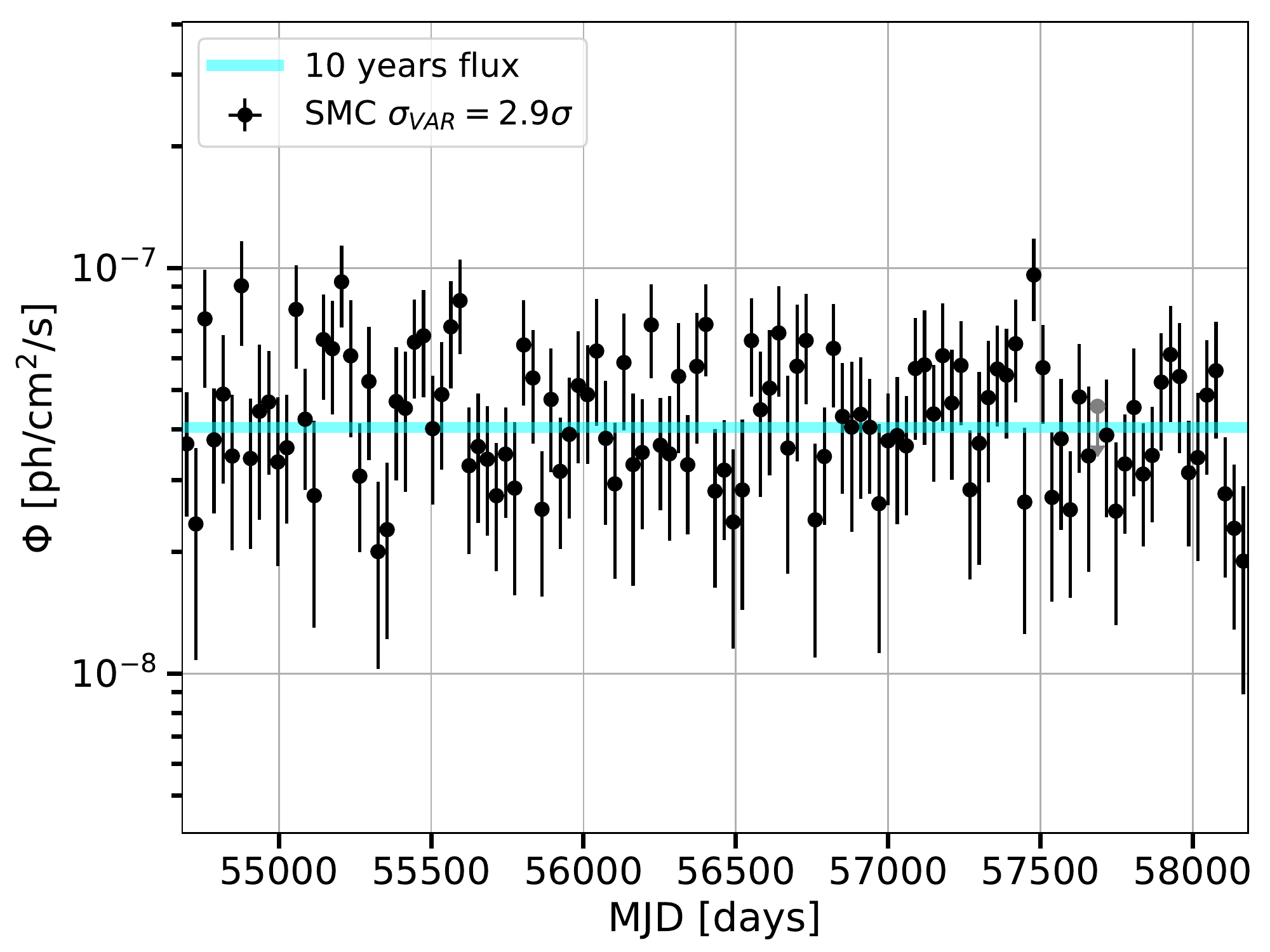}
\includegraphics[width=0.48 \textwidth]{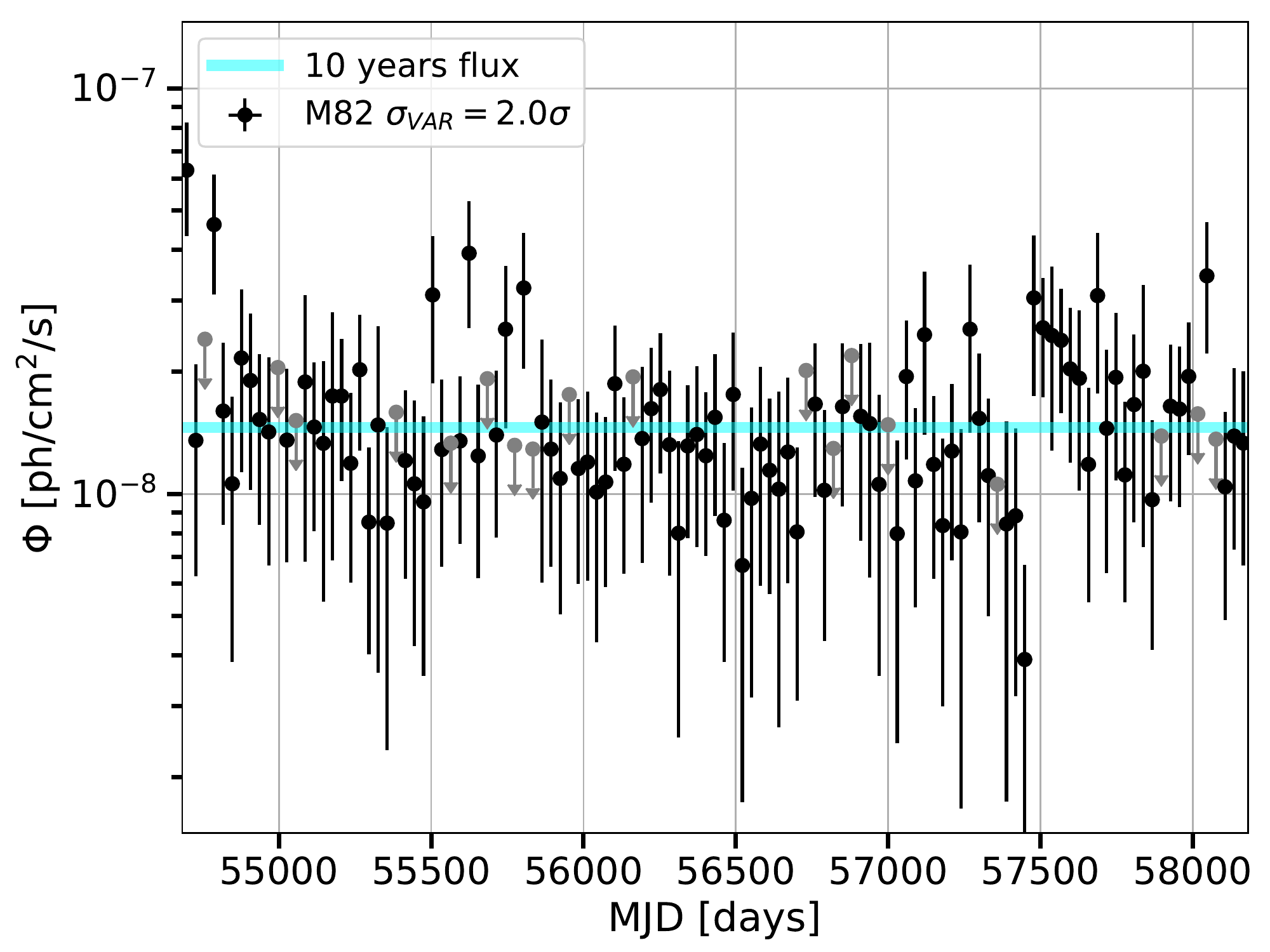}
\includegraphics[width=0.48 \textwidth]{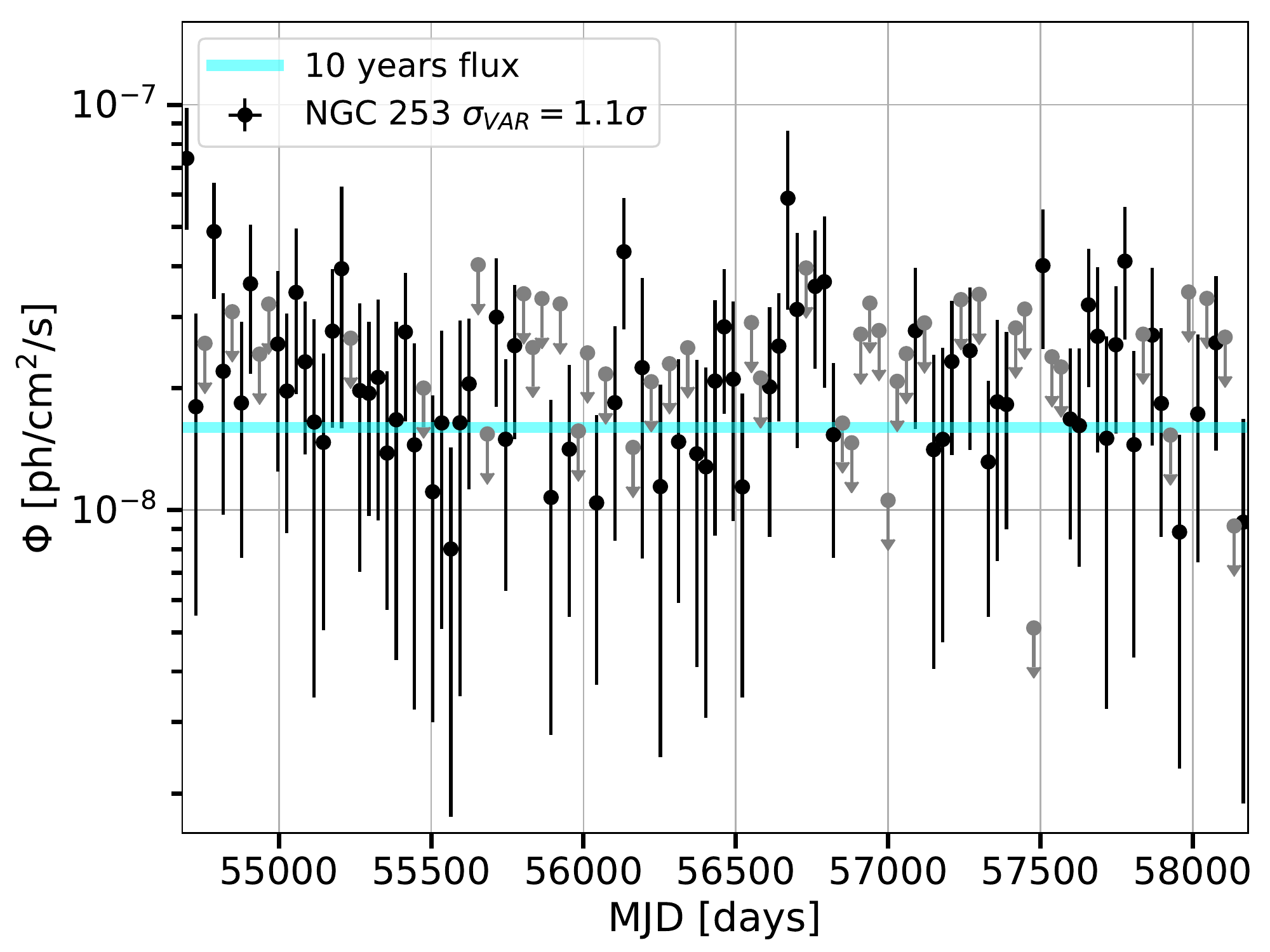}
\includegraphics[width=0.48 \textwidth]{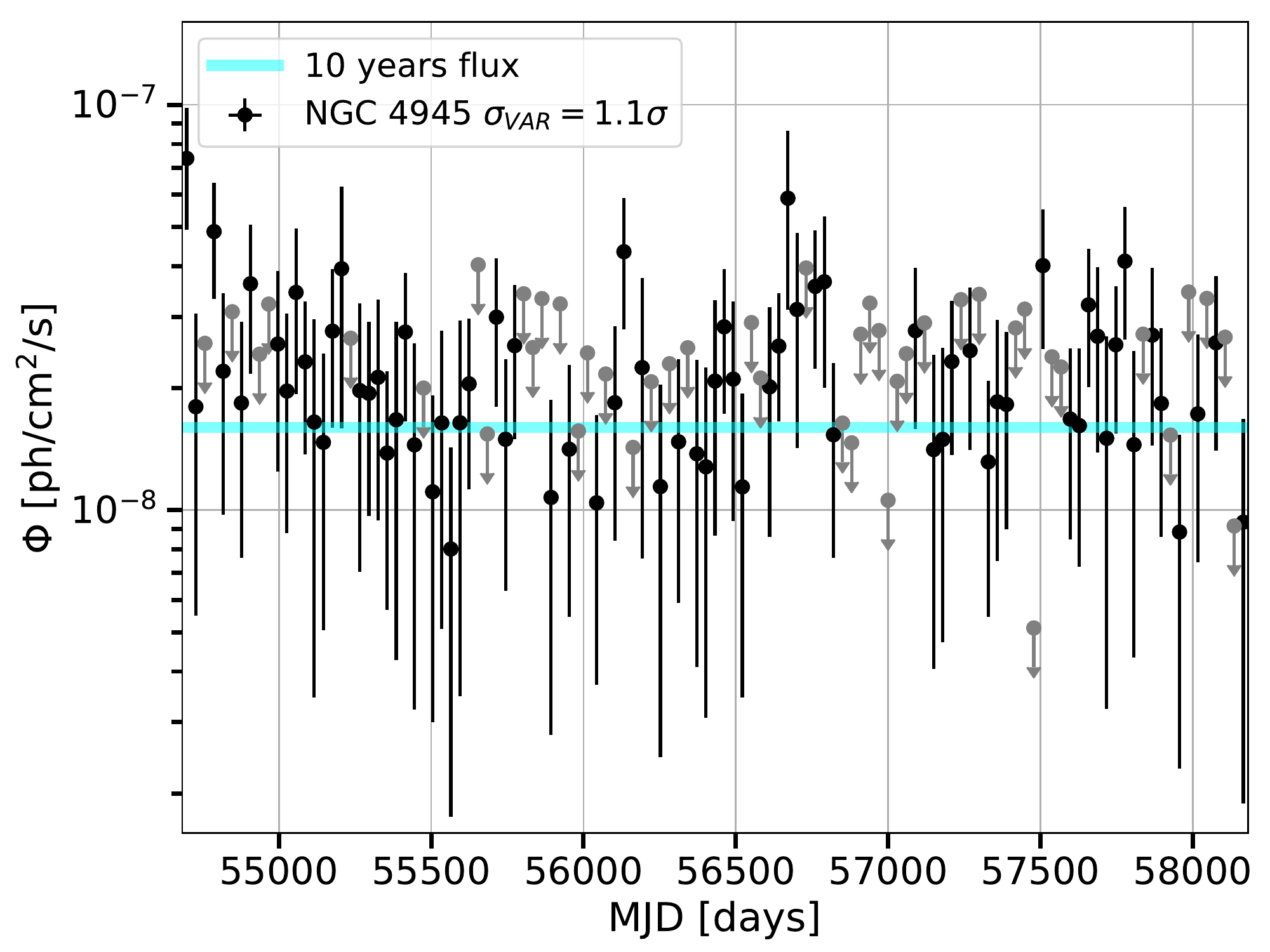}
\includegraphics[width=0.48 \textwidth]{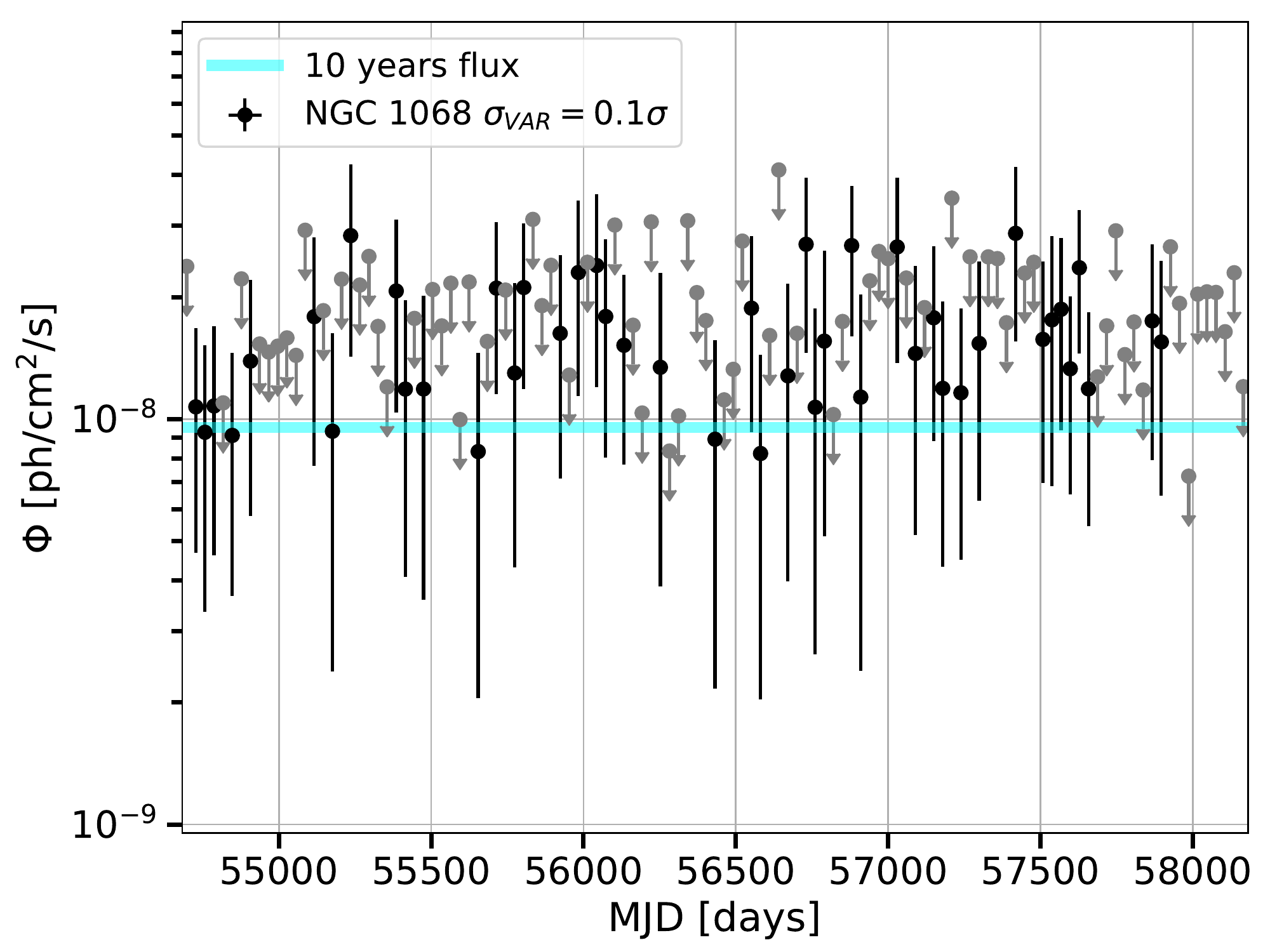}
\caption{Light curves for the SFGs detected in our sample. The black
  data show the flux in each time bin and the cyan lines indicate   the fluxes
  measured in the full dataset. The gray downward arrows are
  1\,$\sigma$ upper limits
  to the source flux {computed when the source $TS<6$}.}
\label{fig:lc1}
\end{figure}

\begin{figure}
\centering
\includegraphics[width=0.48 \textwidth]{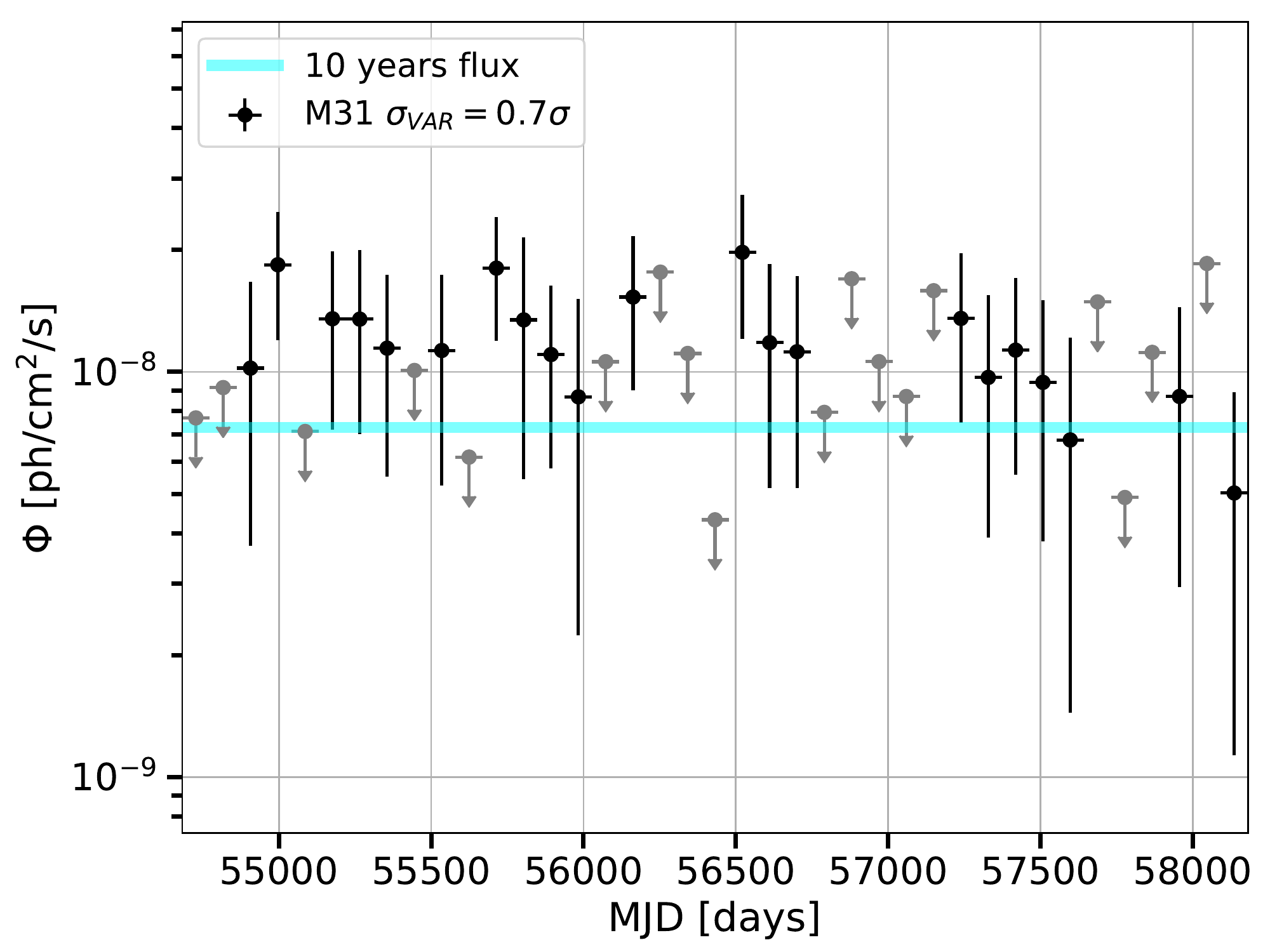}
\includegraphics[width=0.48 \textwidth]{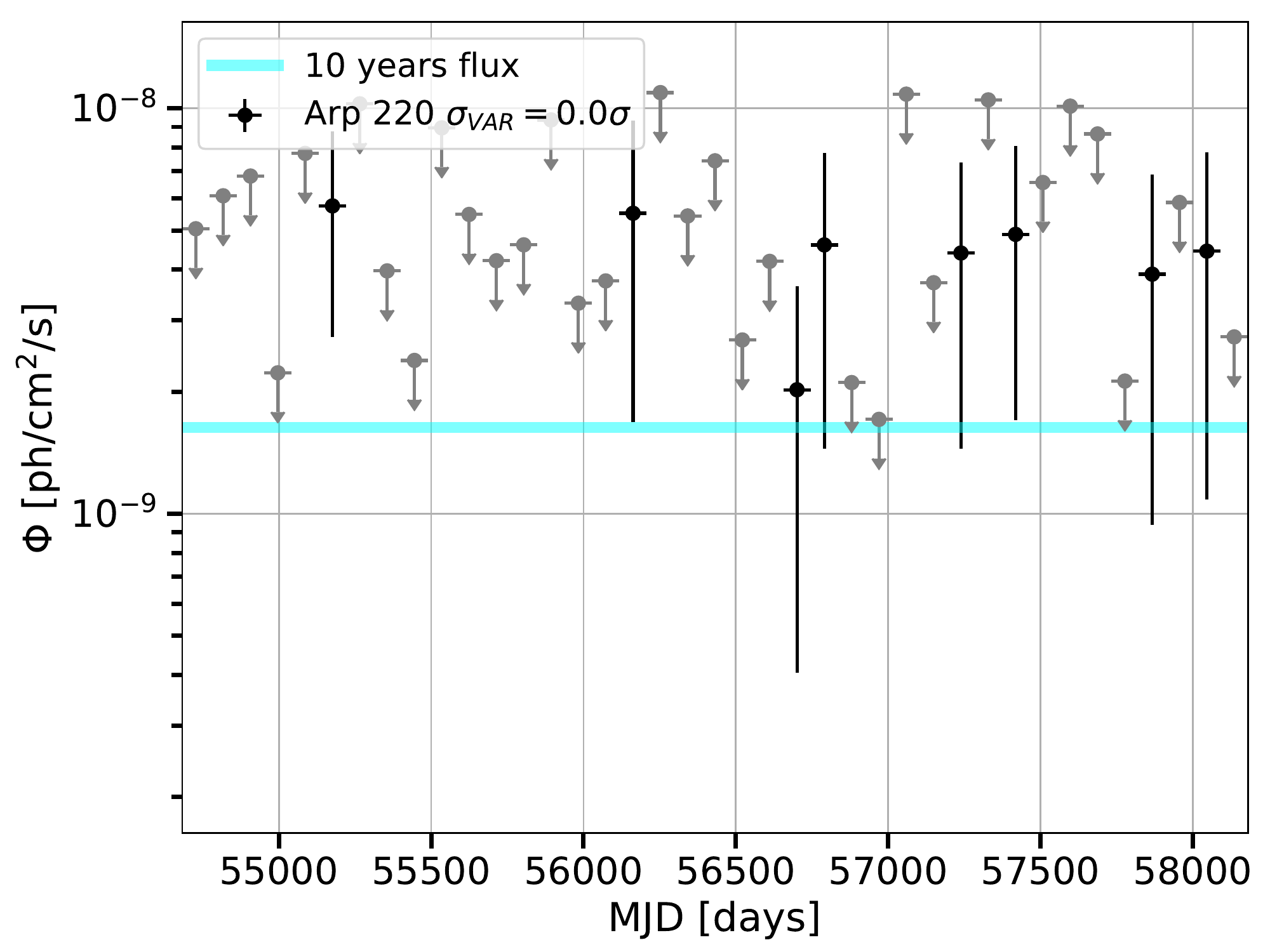}
\includegraphics[width=0.48 \textwidth]{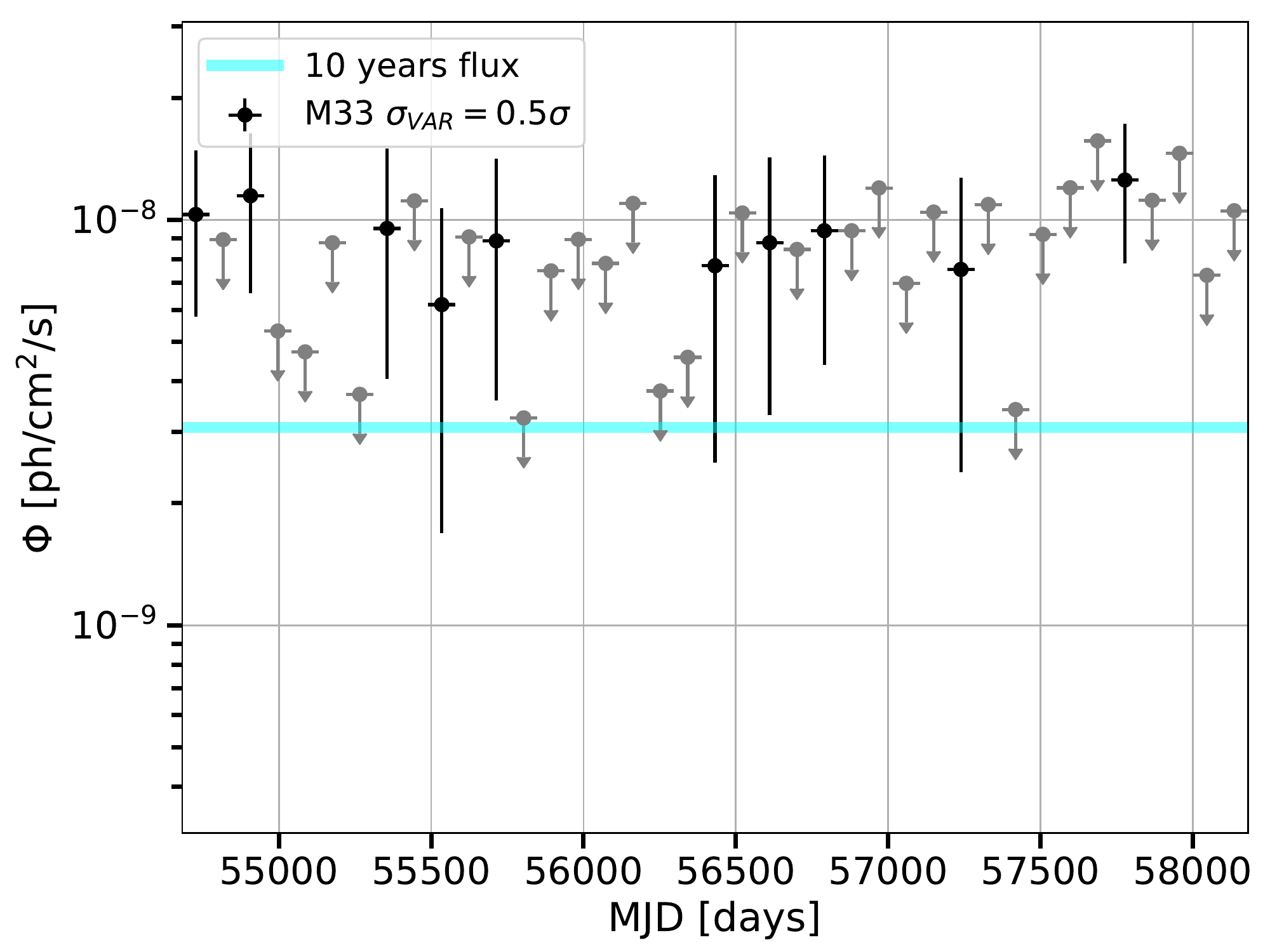}
\includegraphics[width=0.48 \textwidth]{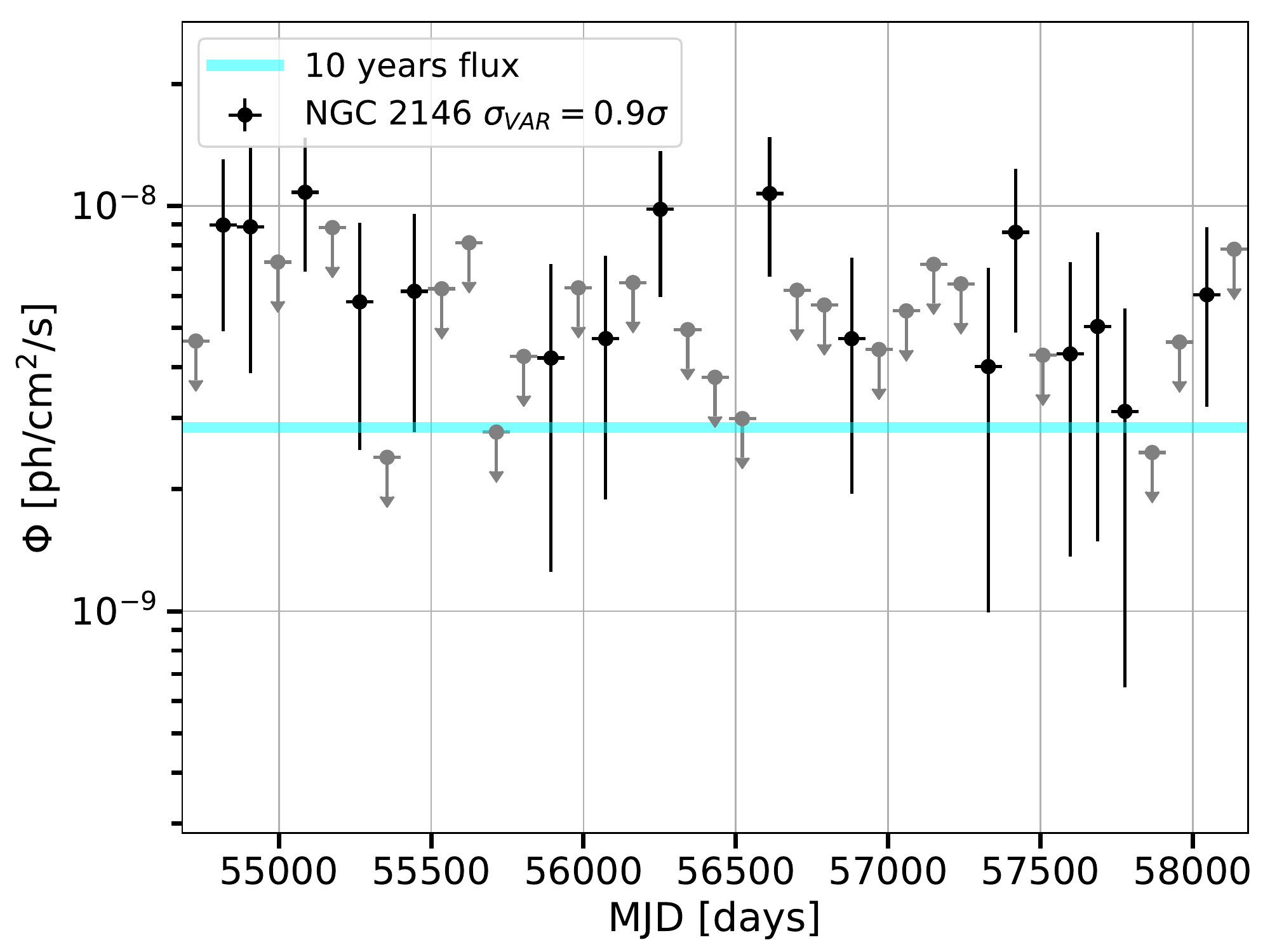}
\includegraphics[width=0.48 \textwidth]{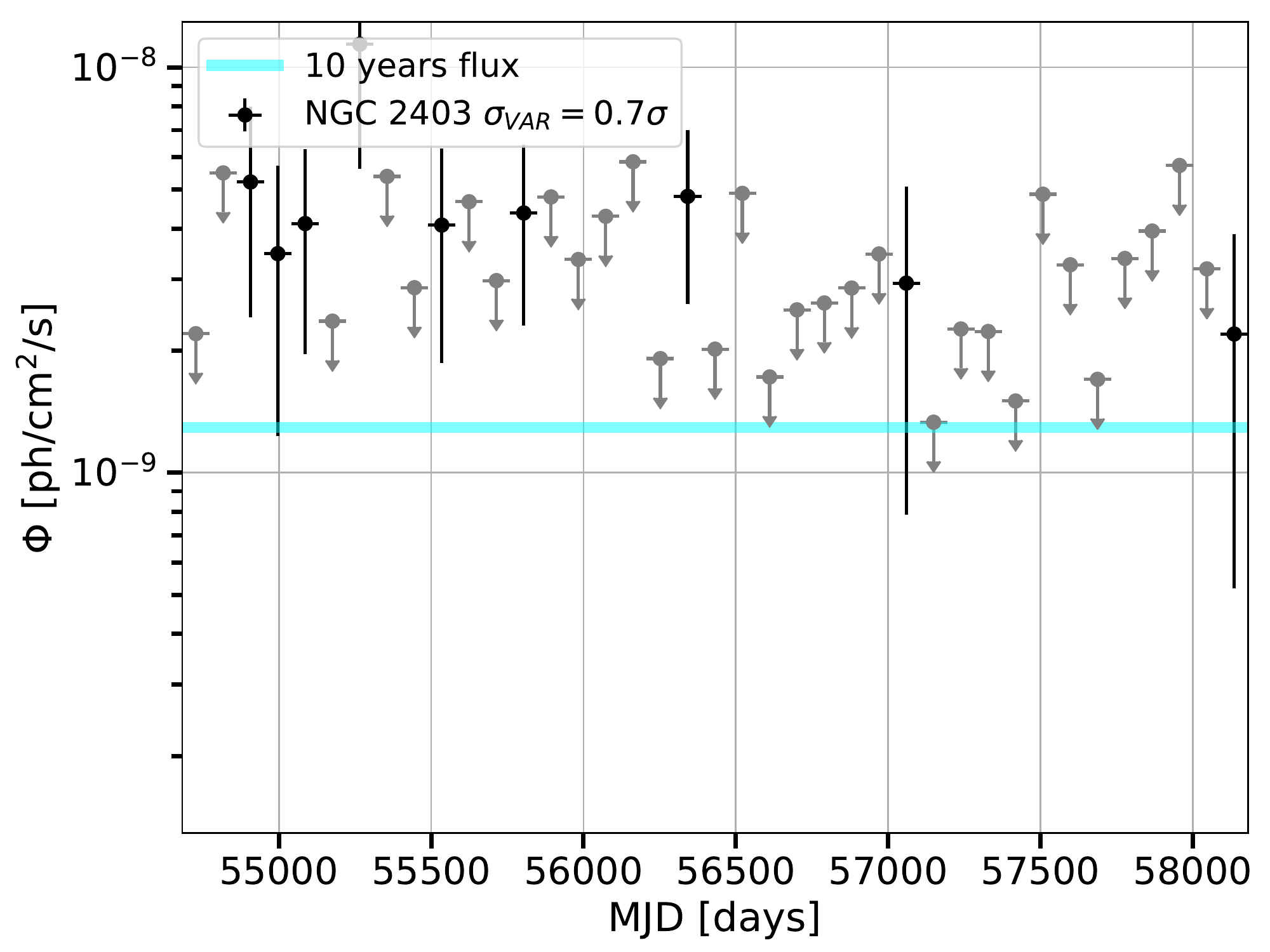}
\includegraphics[width=0.48 \textwidth]{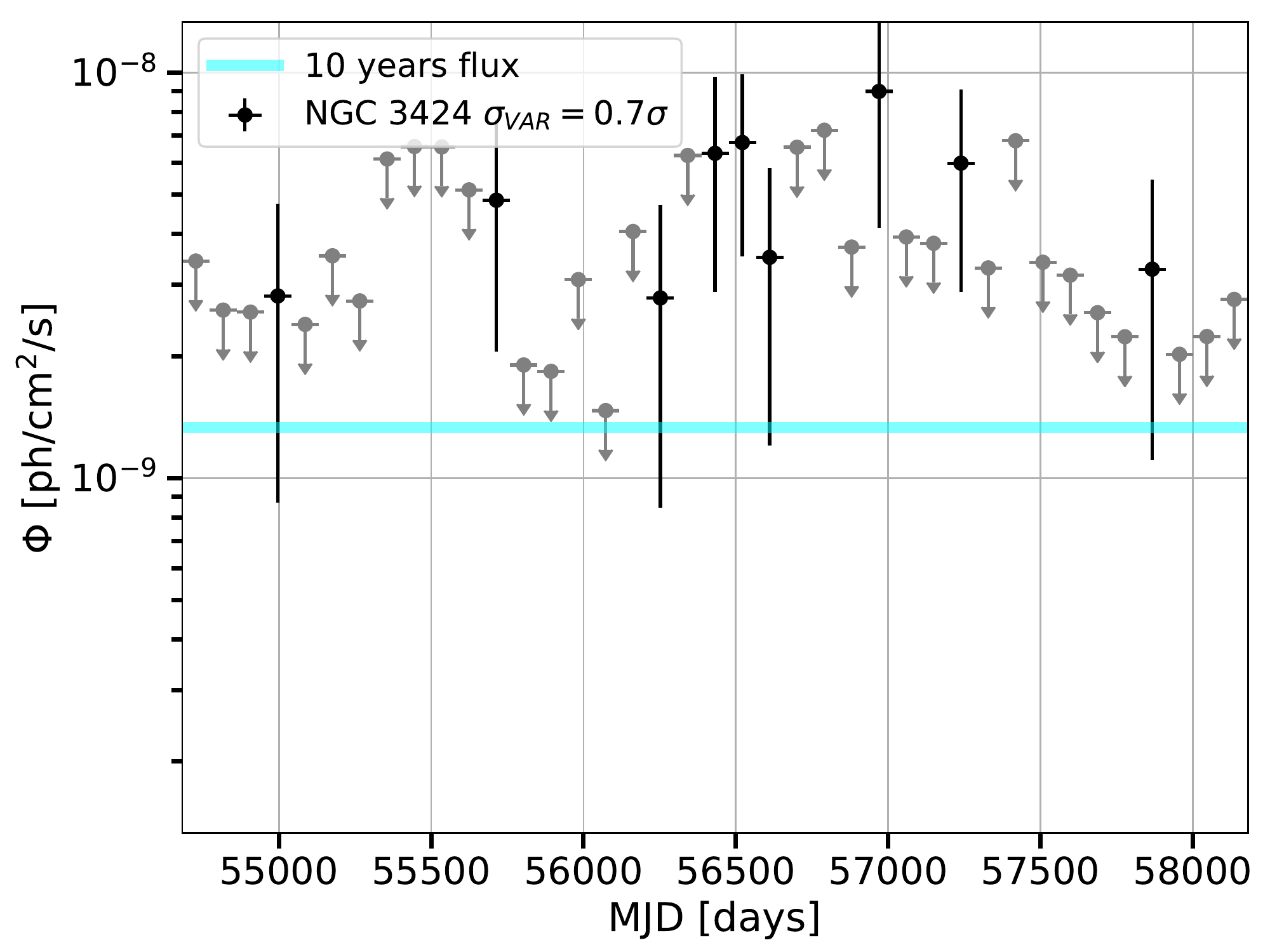}
\caption{Same as in Figure~\ref{fig:lc1}.}
\label{fig:lc2}
\end{figure}

\begin{figure}
\centering
\includegraphics[width=0.48 \textwidth]{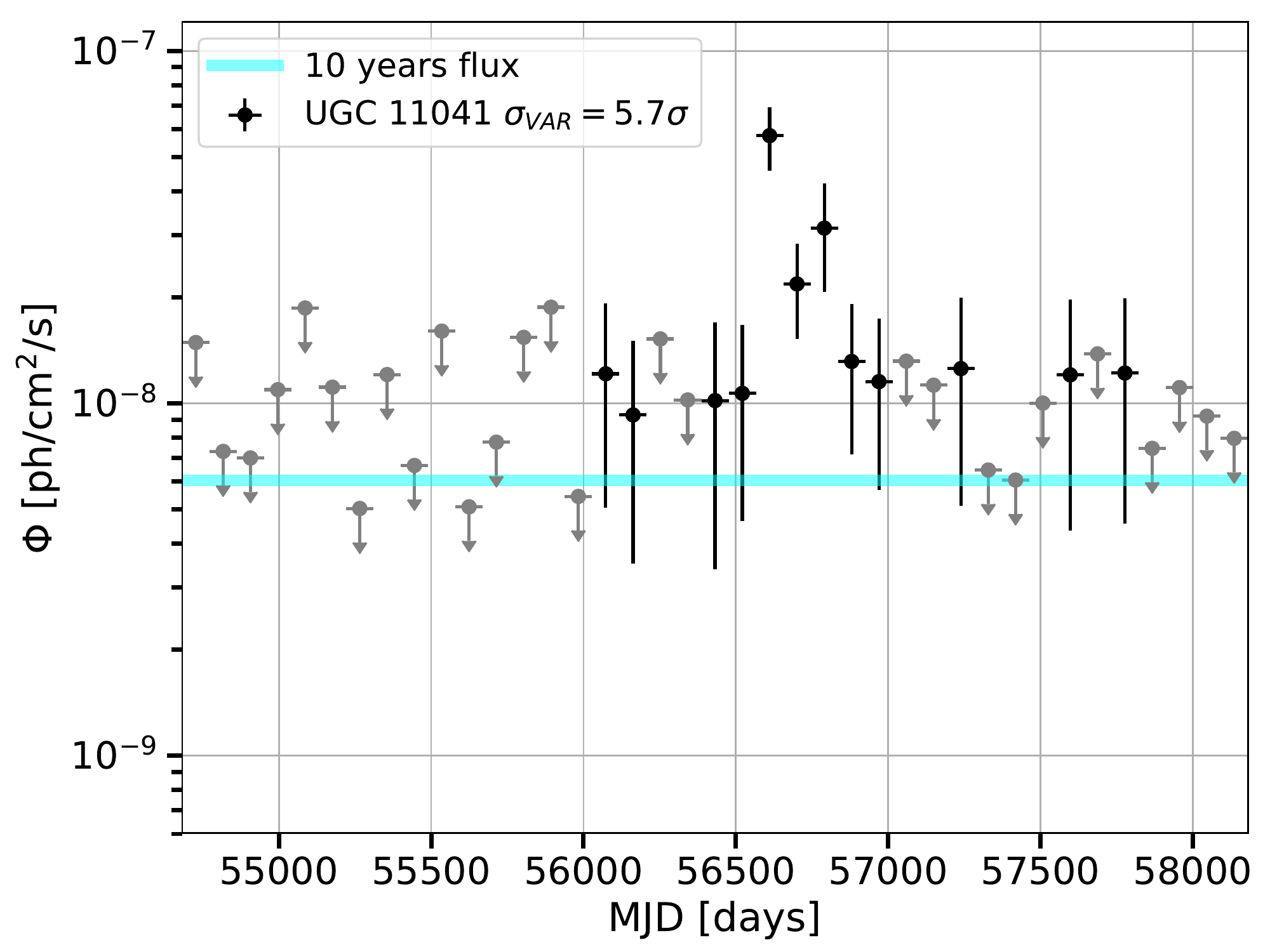}
\includegraphics[width=0.48 \textwidth]{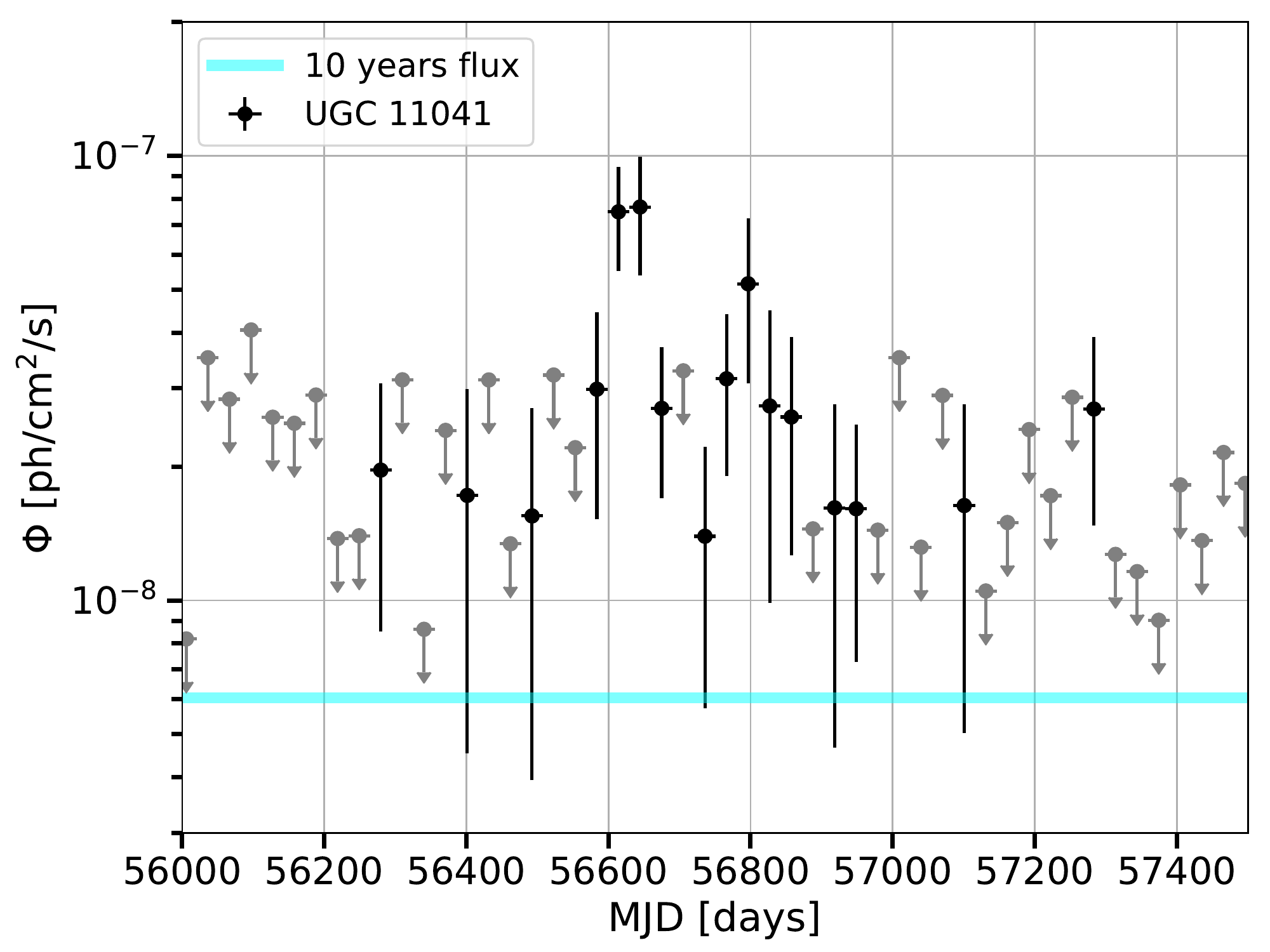}
\caption{Same as in Figure~\ref{fig:lc1}.}
\label{fig:lc3}
\end{figure}

In this section we report the details of the light curve analysis
performed on the sample of SFGs.
We use the function {\tt gta.lightcurve} implemented in {\tt Fermipy}.
This function performs a fit to the ROI independently in each time bin.
For each source {\tt Fermipy} calculates the $TS_{\rm{var}}$, which
follows the definition provided in \cite{2FGL} as: \begin{equation}
     \label{rho_gamma3}    
      TS_{\rm{var}}  = \sum_i   [\log{L_i(F_i)}-\log{L_i(F_{\rm{const}})}],
\end{equation}
where $\log{L_i(F_i)}$ is the log-likelihood of the fit with the SED parameters 
of the source of interest free to vary in each  $i$-th time bin, and $\log{L_i(F_{\rm{const}})}$ 
is the log-likelihood for the fit at the same time bin with the SED parameters fixed 
to provide the flux found in the entire time range.
We add, as done in  {\it Fermi}-LAT catalogs, a $2\%$ systematic uncertainty
to account for systematic errors in the calculation
of the source exposure, resulting from small inaccuracies in
the dependence of the IRFs on the source viewing angle,
coupled with changes in the observing profile as the orbit of
the spacecraft precesses. 
The results for the variability are reported in
Table~\ref{tab:seddetected} as significance for the variability
$\sigma_{\rm{var}}$, calculated by converting  $TS_{\rm{var}}$ to a
significance by adopting the $\chi^2$ distribution 
with 118 degrees of freedom for the six
brightest sources and 38 degrees of freedom for the others. The lightcurves of all
sources are shown in Figs.~\ref{fig:lc1}, \ref{fig:lc2},  and \ref{fig:lc3}. 

The only significantly variable object is UGC 11041 with $\sigma_{\rm{var}}=5.7$.
This source had a bright flare during the period 56500--56800\,MJD. 
In this time range the source is detected with a $TS=76$ and has a
best-fit position of $RA(J2000) = 268.72^{\circ} \pm 0.07^{\circ}$ and
$DEC(J2000)=34.76^{\circ} \pm 0.07^{\circ}$, which is fully compatible with
the galaxy's known position.
We also run the light-curve analysis for this source with a finer time
bin of one month. This is shown in the right-hand panel of Figure~\ref{fig:lc3} which demonstrates that the time scale for the flare is of about two months.
The position of this source is coincident with the radio source NVSS J175451+344633.
This indicates that the observed $\gamma$-ray emission is not due to
star-formation activity, but rather points to
the emission of a relativistic jet.

We test the accuracy of our lightcurve extraction method by
analyzing pulsars, which are the least variable $\gamma$-ray sources
on long timescales.
We select pulsars from  the 3FGL \citep{3FGL} with  variability
indices smaller than 55, detected 
at $|b|>4^{\circ}$, at least $20^{\circ}$ away from the Galactic
center, and with a statistical significance greater than $8$\,$\sigma$.
This leaves us with 50 pulsars on which we run our variability analysis using one time interval per month.
We find that these pulsars have an average $TS_{\rm{var}}$ of 110, which implies
$\sim 0.40$\,$\sigma$ significance for the variability.
The mean and standard deviation for the $TS_{\rm{var}}$ distribution
are 110 and 45, respectively.
Therefore, all the results for our SFGs, except for UGC 11041, are consistent with the results found for our sample of pulsars.
The most-variable pulsars in our sample are the following: PSR J1836+5925 ($\sigma_{\rm{var}}=6.2\sigma$), PSR J1231$-$1411 ($\sigma_{\rm{var}}=5.6\sigma$), PSR J1057$-$5226 ($\sigma_{\rm{var}}=4.8\sigma$), PSR  J0218+4232($\sigma_{\rm{var}}=4.7\sigma$),  PSR J1311$-$3430 ($\sigma_{\rm{var}}=4.2\sigma$) and PSR J0357+3205 ($\sigma_{\rm{var}}=4.0\sigma$).



\begin{figure}
\centering
\includegraphics[width=0.48 \textwidth]{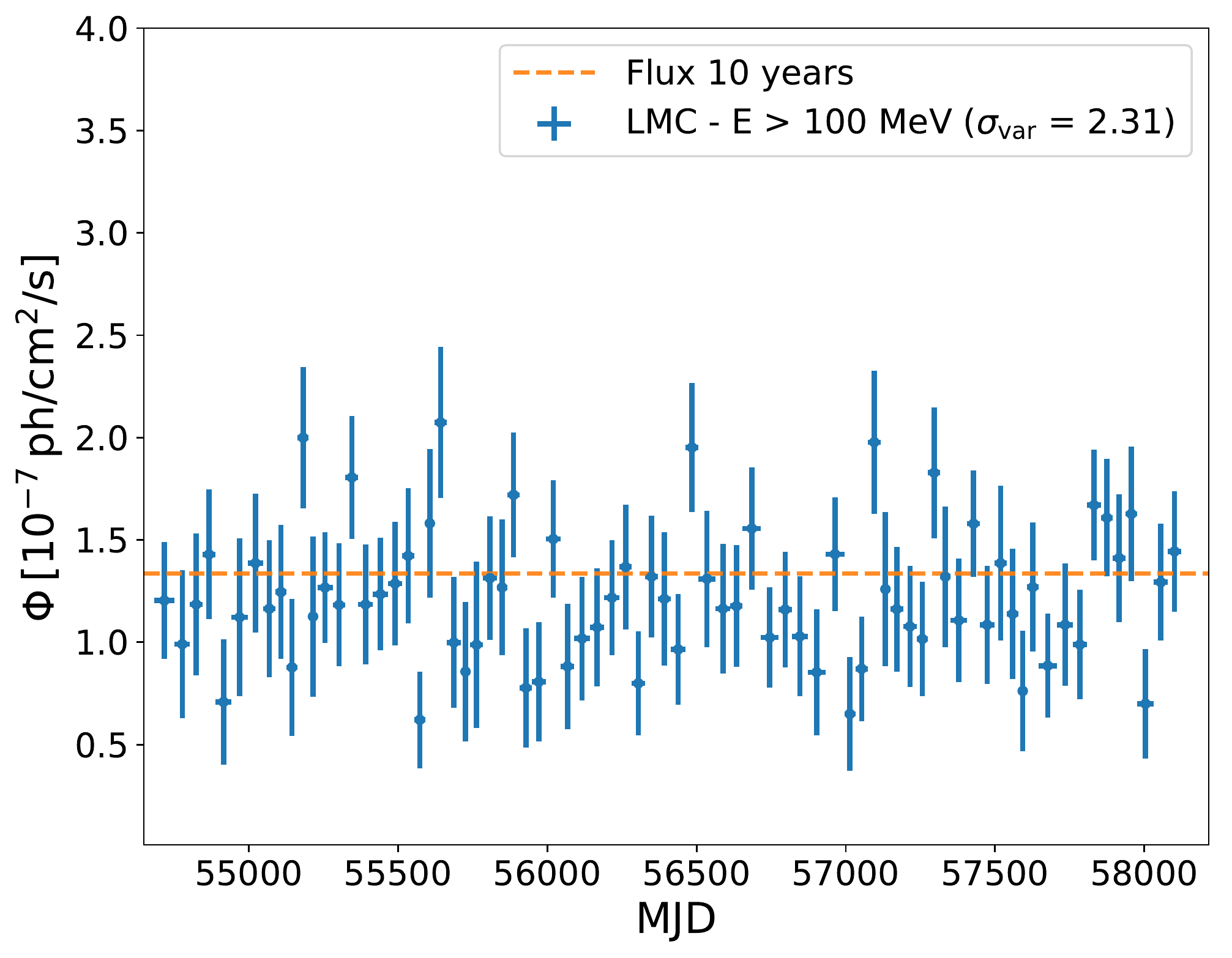}
\includegraphics[width=0.48\textwidth]{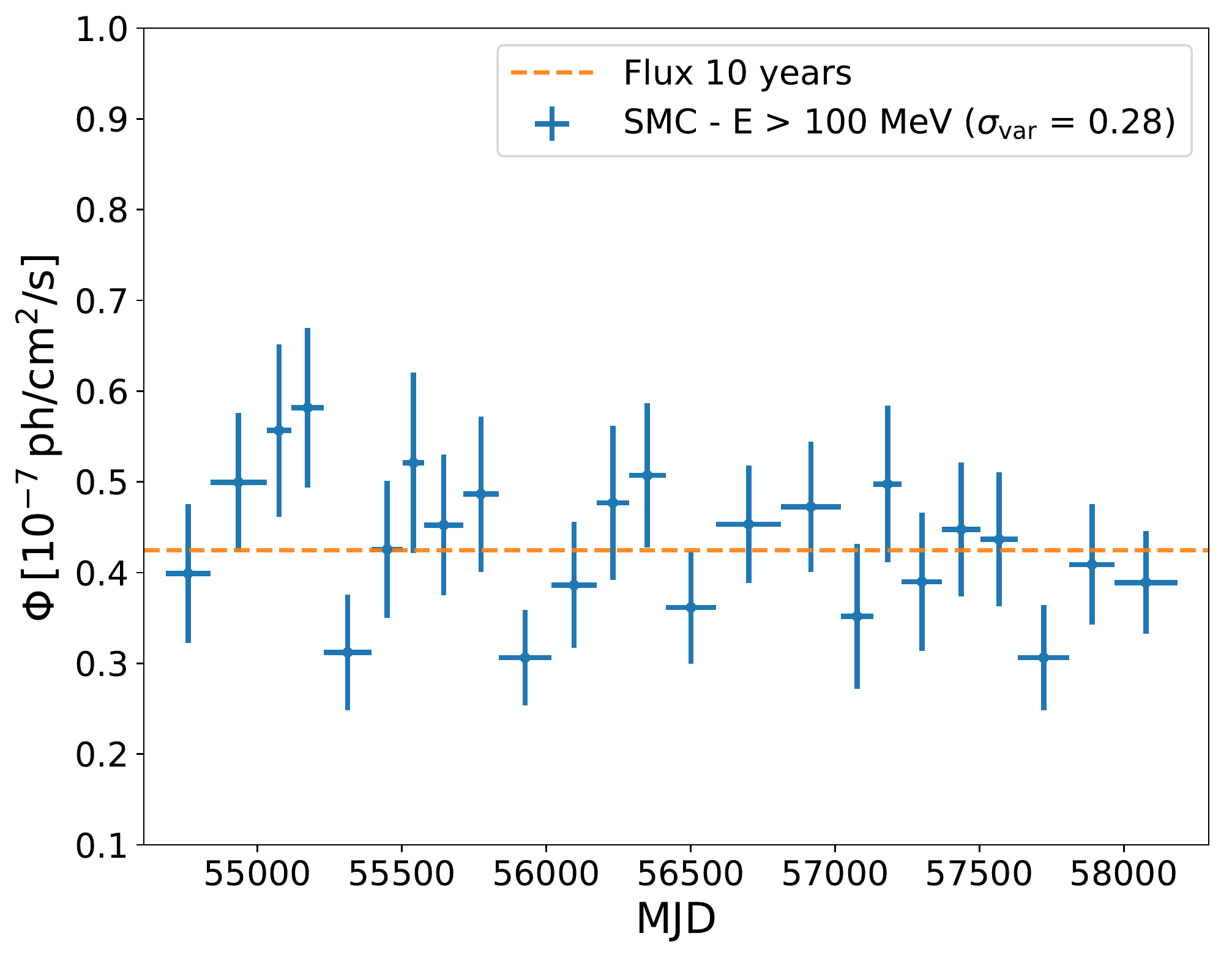}
\includegraphics[width=0.48 \textwidth]{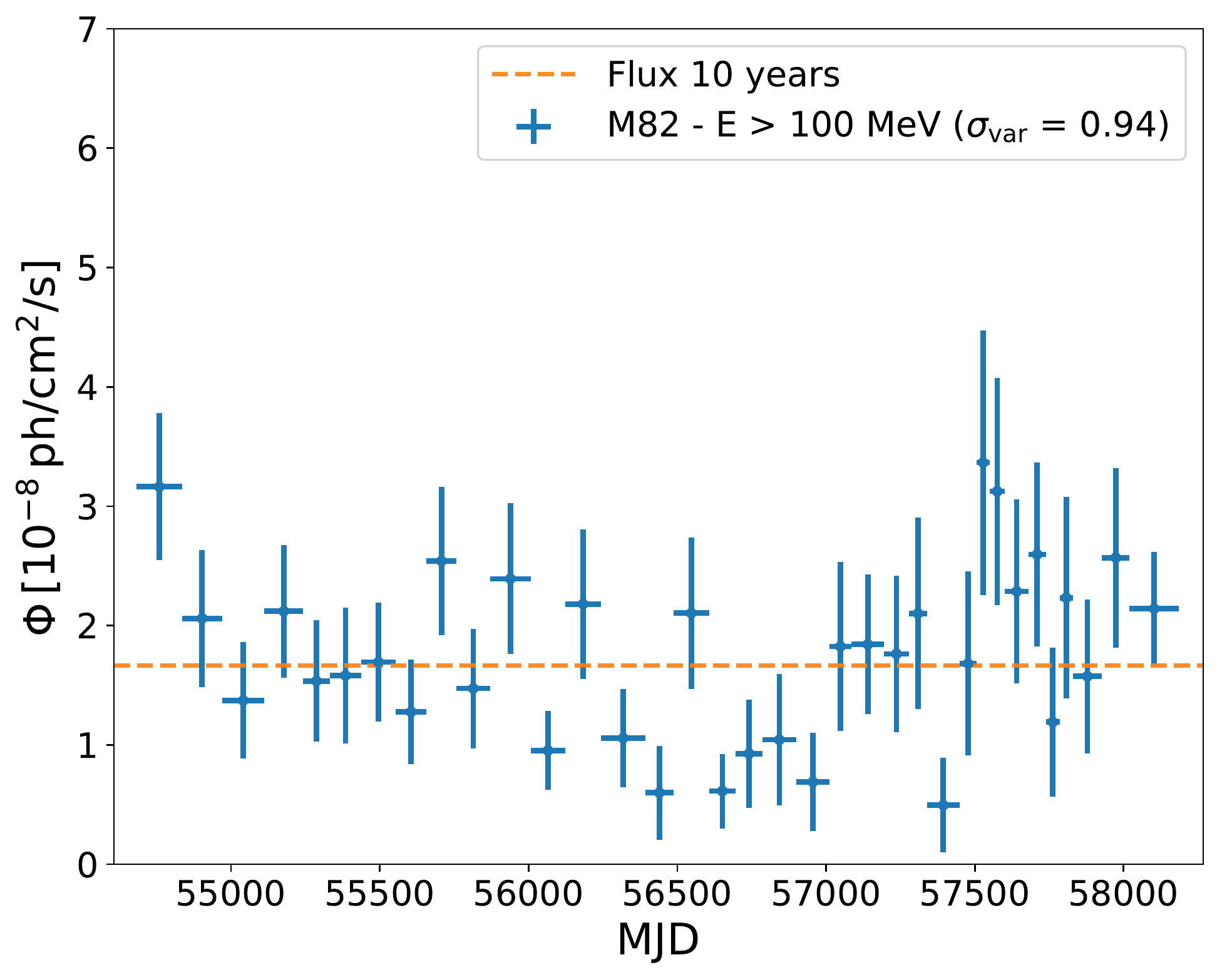}
\includegraphics[width=0.48\textwidth]{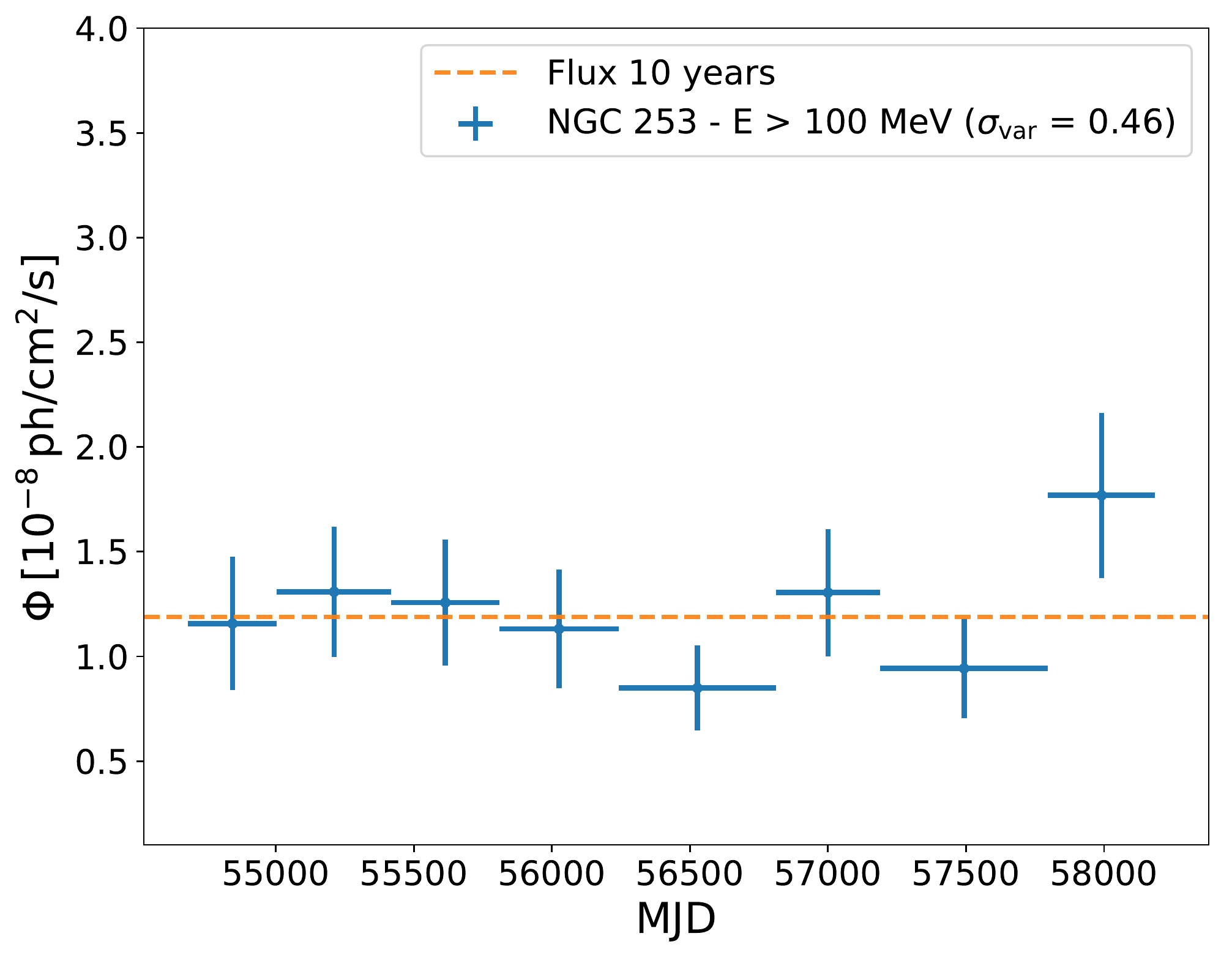}
\caption{Constant signal-to-noise ratio light curves of the brightest galaxies.}
\label{fig:lc_ab1}
\end{figure}

\begin{figure}
\centering
\includegraphics[width=0.48 \textwidth]{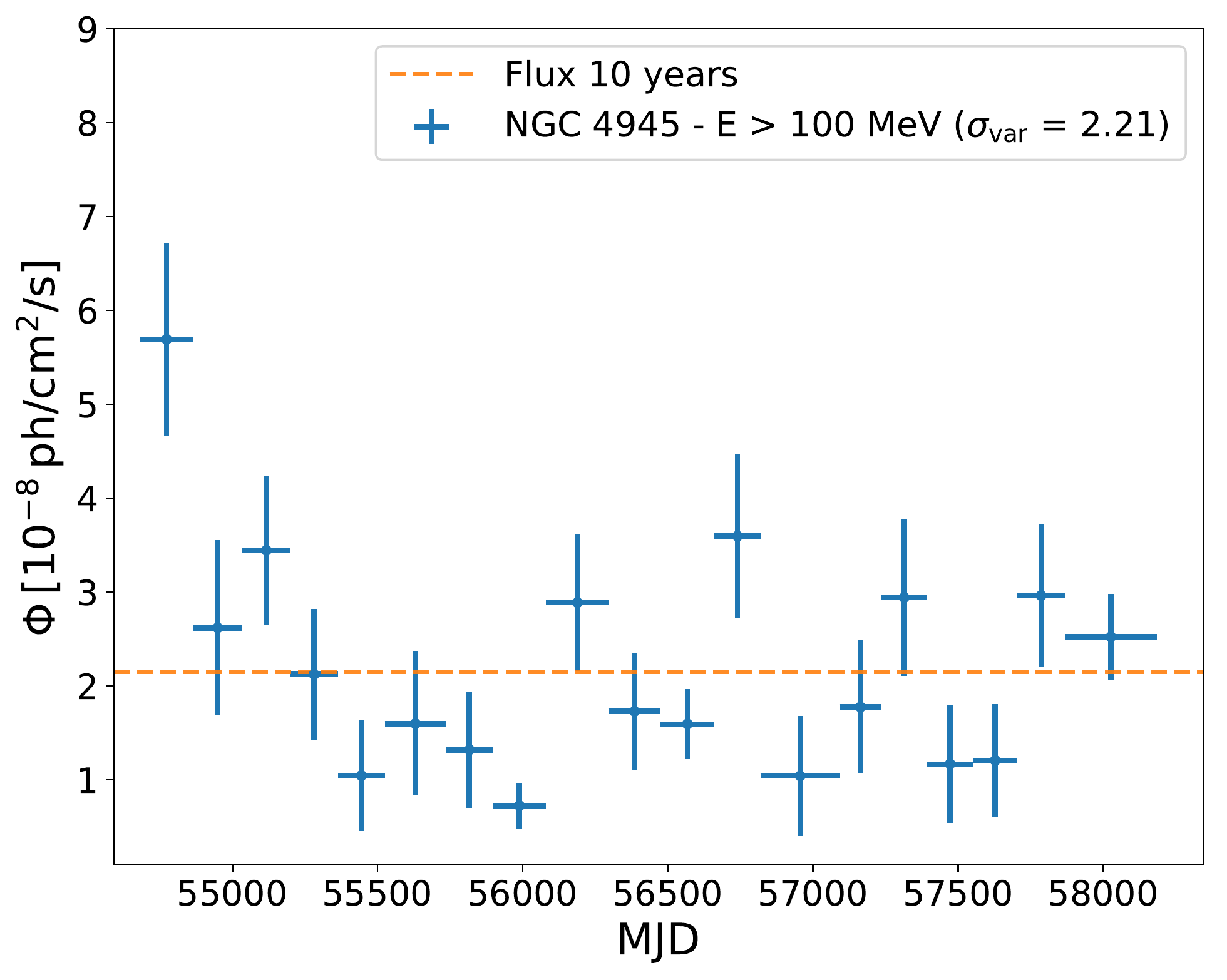}
\includegraphics[width=0.48\textwidth]{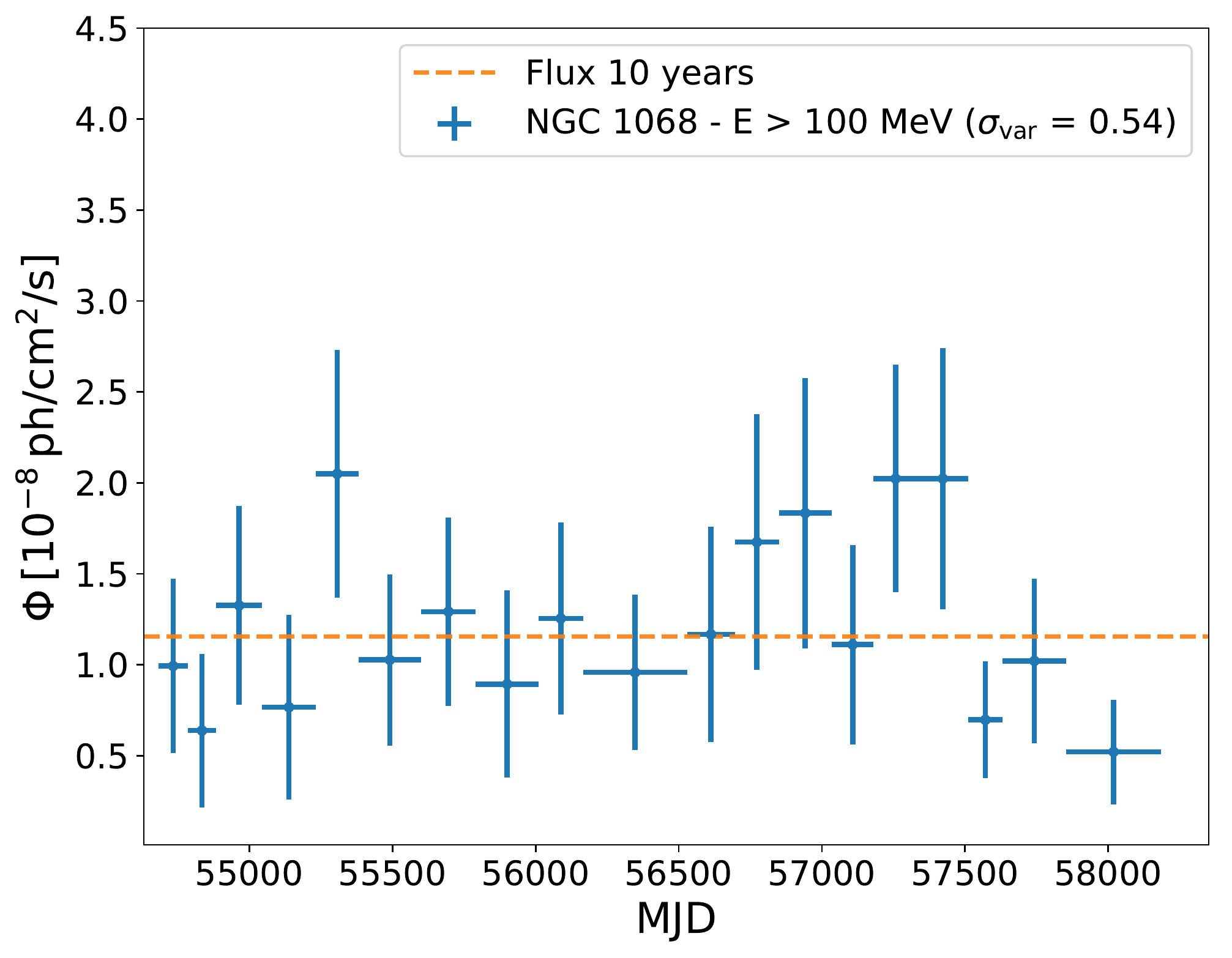}
\includegraphics[width=0.48 \textwidth]{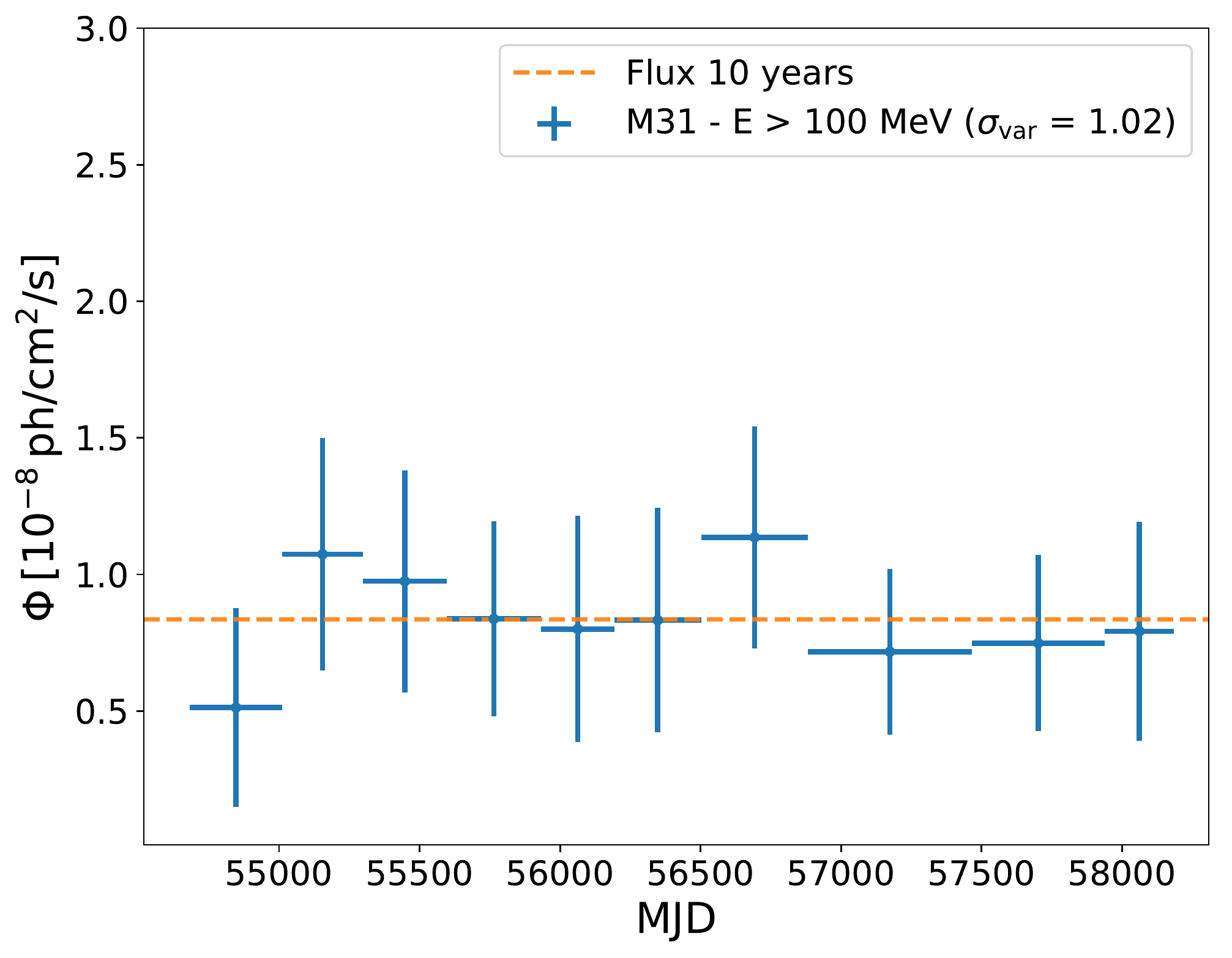}
\caption{Same as in Figure~\ref{fig:lc_ab1}.}
\label{fig:lc_ab2}
\end{figure}

Finally, we produce lightcurves for the brightest SFGs
using the adaptive binning method of \cite{lott2012}.
In this method, the time bins have been optimized to have a
fractional uncertainty of 15\,\% for all sources except NGC~1068 and
M31, for which a fractional uncertainty of 20\,\% was used.
The lightcurves, reported in Figures~\ref{fig:lc_ab1} and \ref{fig:lc_ab2},
confirm the overall result of the lack of variability for our sample
of galaxies.

\begin{figure}
\centering
\includegraphics[width=0.50 \textwidth]{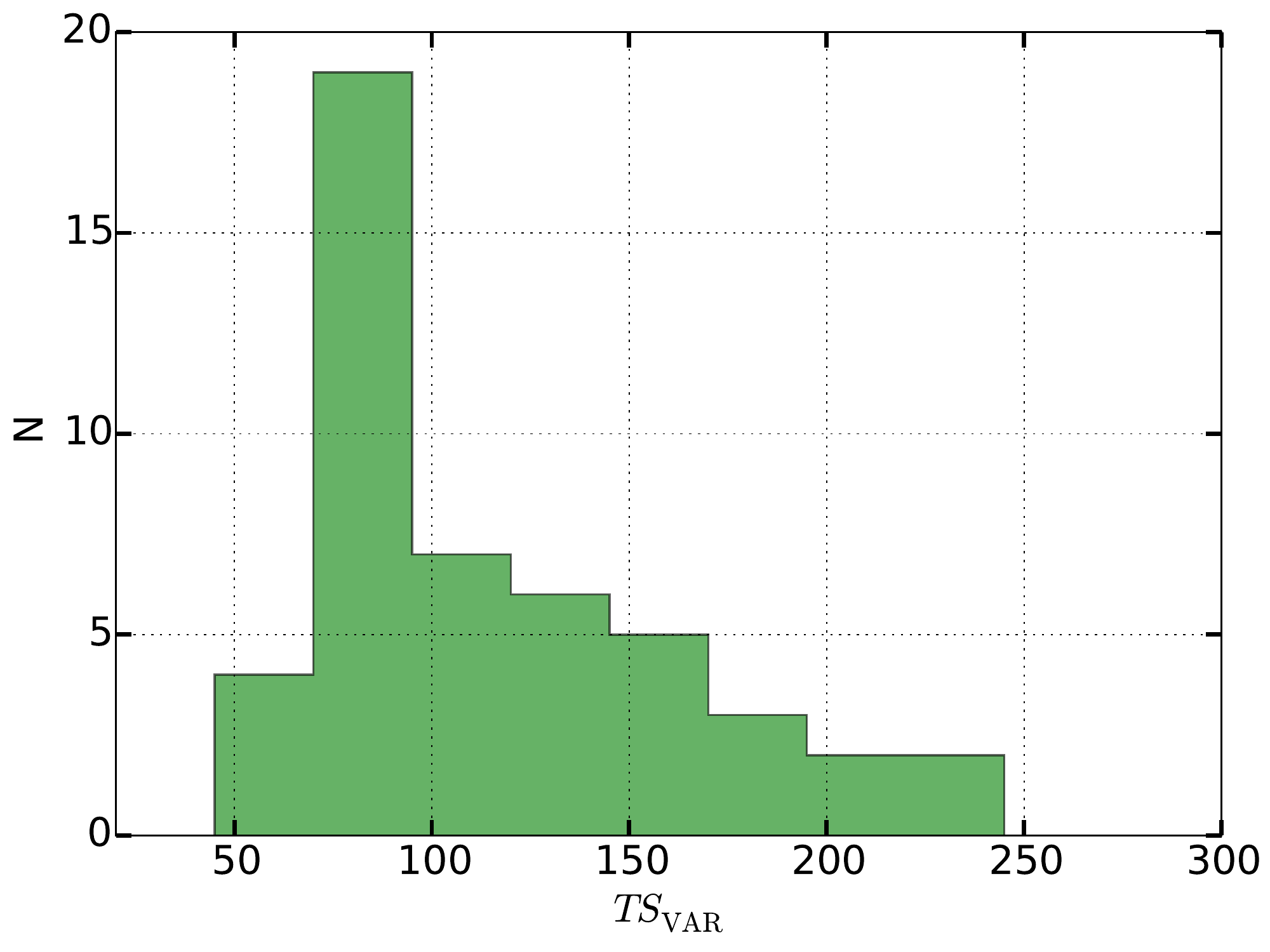}
\caption{Histogram of the $TS_{\rm{var}}$ for the 50 selected pulsars in our sample.} 
\label{fig:histoPSRs}
\end{figure}

%
%
\section{Validation of The Stacking Technique}

%
%

\begin{figure*}[ht!]
  \begin{center}
  \begin{tabular}{c}
  	 \includegraphics[scale=0.5,clip=true,trim=0 0 0 0]{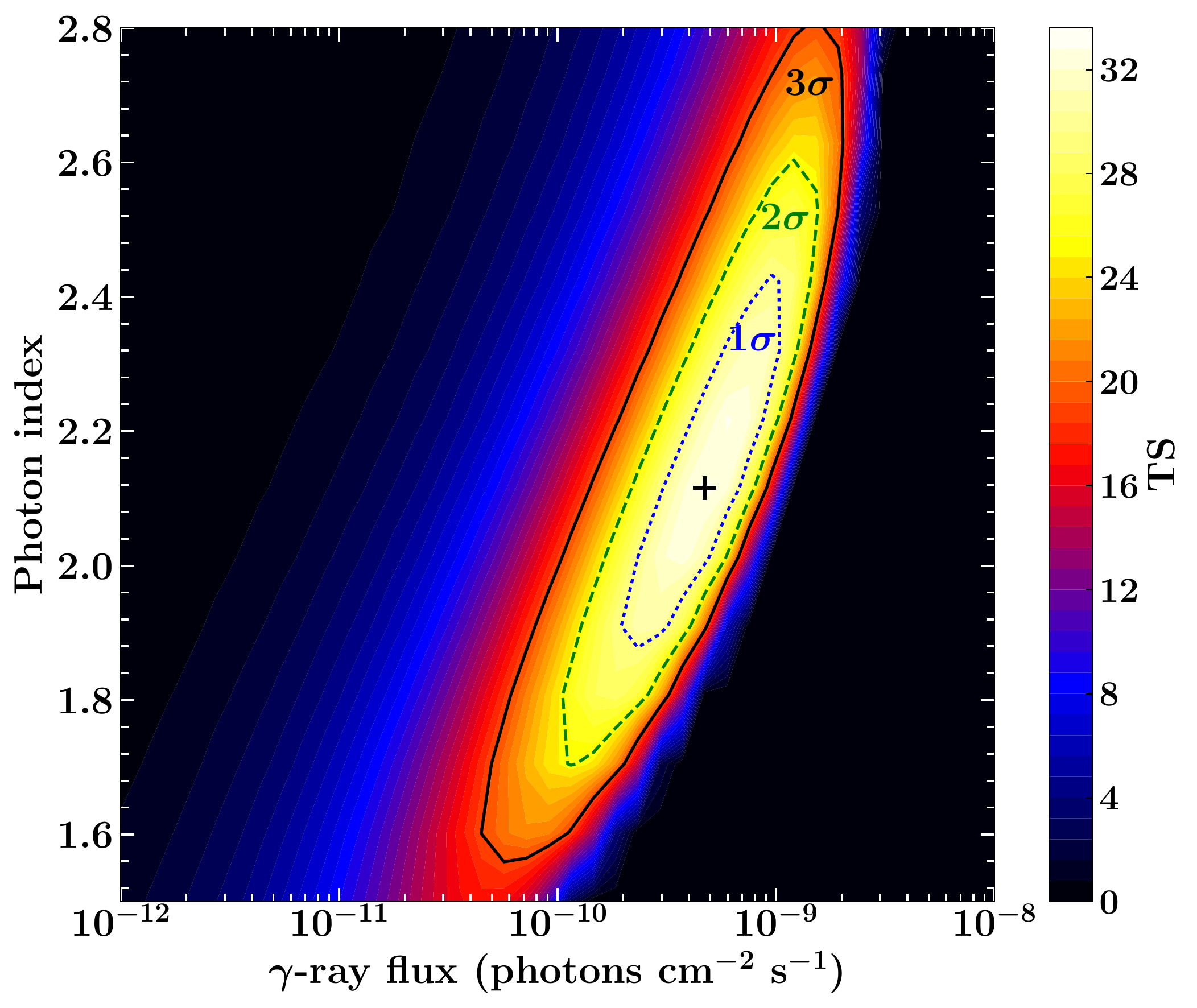} \\
\end{tabular}
  \end{center}
  \caption{Stack of 20 simulated sources with average photon index of
    2.2$\pm0.2$ and average (100\,MeV -- 800\,GeV) flux of
    $6.1(\pm1.8)\times 10^{-10}$\,ph cm$^{-2}$ s$^{-1}$ 
\label{fig:sim}}
\end{figure*}

\label{sec:sim}
In order to be able to interpret correctly the results of the stacking
analysis we performed a test where
we simulated 10\,years of Pass 8 data for 20 sources at $|b|>10^{^\circ}$, whose spectra are
 modeled using a power law. The power-law indices are extracted from
a Gaussian distribution with a mean of 2.2 and a dispersion of 0.2, while
the fluxes are extracted from a power-law distribution with index $-$2.5
having an average of 6.41$\times10^{-10}$\,ph cm$^{-2}$ s$^{-1}$.
The simulations include the
diffuse (Galactic and isotropic) emission as well as all the point
sources from the FL8Y.
A stacking analysis was performed for the
simulated dataset as described in $\S$~\ref{sec:stacking}, and the
results are presented in Figure~\ref{fig:sim}. The average emission of
these simulated sources is detected with a $TS$$\approx$32. The best-fit
photon index and flux are 2.12$\pm0.12$ and
5.0$\pm2.0\times10^{-10}$\,ph cm$^{-2}$ s$^{-1}$, respectively, which is in agreement with
the average values  of the simulated sources.
From this analysis we conclude that the stacking analysis provides a measurement of the
average properties (in this case photon index and flux) of the
unresolved population.

\bibliographystyle{apj}
\bibliography{sfg}

\end{document}